\documentclass[lettersize,journal]{IEEEtran}
\usepackage{amsmath,amsfonts}
\usepackage{algorithmic}
\usepackage{algorithm}
\usepackage{array}
\usepackage{textcomp}
\usepackage{stfloats}
\usepackage{url}
\usepackage{verbatim}
\usepackage{graphicx}
\usepackage{cite}

\usepackage{upgreek}

\usepackage[colorlinks,urlcolor=blue,linkcolor=blue,citecolor=blue]{hyperref}
\usepackage{color,array,amsthm}

\usepackage{stfloats}
\usepackage{enumerate}
\usepackage{amsmath}
\usepackage{booktabs}
\usepackage{float}
\usepackage{bm}
\usepackage{stfloats}
\usepackage{makecell}
\usepackage{epstopdf}
\usepackage{multirow}
\usepackage{threeparttable}
\usepackage{pifont}
\usepackage{tabularx}

\usepackage{graphicx}
\usepackage[caption=false,font=scriptsize,labelfont=rm,textfont=rm]{subfig}

\usepackage[T1]{fontenc}
\usepackage[utf8]{inputenc}

\usepackage{enumitem}

\begin{document}

\title{MIMO OFDM-Enabled ISAC for Low-Altitude Non-Cooperative UAV Surveillance: A Survey}

\author{Shiyu Bai$^{\dagger}$,~\IEEEmembership{Member,~IEEE,} Sijia Li$^{\dagger}$,~\IEEEmembership{Graduate Student Member,~IEEE,} Cunyi Yin,~\IEEEmembership{Member,~IEEE,} Wenqiu Qu, Li-Ta Hsu,~\IEEEmembership{Senior Member,~IEEE}, Yuanwei Liu,~\IEEEmembership{Fellow,~IEEE}, Wen-Hua Chen,~\IEEEmembership{Fellow,~IEEE}
        % <-this % stops a space
\thanks{$^{\dagger}$Shiyu Bai and Sijia Li contributed equally to this work.}
\thanks{Shiyu Bai is with the Hong Kong Polytechnic University, Hong Kong, China (e-mail: \href{mailto:shiyu.bai@polyu.edu.hk}{shiyu.bai@polyu.edu.hk}). Sijia Li is with the Hong Kong Polytechnic University, Hong Kong, China (e-mail: \href{mailto:sijia-franz.li@connect.polyu.hk}{sijia-franz.li@connect.polyu.hk}). Cunyi Yin is with the Hong Kong Polytechnic University, Hong Kong, China (e-mail: \href{mailto:cunyi.yin@polyu.edu.hk}{cunyi.yin@polyu.edu.hk}). Li-Ta Hsu is with the Hong Kong Polytechnic University, Hong Kong, China (e-mail: \href{mailto:lt.hsu@polyu.edu.hk}{lt.hsu@polyu.edu.hk}). Wen-Hua Chen is with the Hong Kong Polytechnic University, Hong Kong, China (e-mail: \href{mailto:wenhua.chen@polyu.edu.hk}{wenhua.chen@polyu.edu.hk}).}
\thanks{Yuanwei Liu is with The University of Hong Kong, Hong Kong, China (e-mail: \href{mailto:yuanwei@hku.hk}{yuanwei@hku.hk}).}
\thanks{Wenqiu Qu is with the Beihang University, Beijing, China (e-mail: \href{mailto:quwq17@buaa.edu.cn}{quwq17@buaa.edu.cn}).}
\thanks{\itshape (Corresponding author: Wen-Hua Chen).}}

\markboth{Journal of \LaTeX\ Class Files,~Vol.~14, No.~8, August~2021}%
{Shell \MakeLowercase{\textit{et al.}}: A Sample Article Using IEEEtran.cls for IEEE Journals}

\maketitle

\begin{abstract}
The widespread use of unmanned aerial vehicles (UAVs) in low-altitude airspace has raised significant safety and security concerns, motivating the development of reliable non-cooperative UAV surveillance technologies. Integrated sensing and communication (ISAC), enabled by multiple-input multiple-output (MIMO) architectures and orthogonal frequency-division multiplexing (OFDM) waveforms, has emerged as a promising paradigm for leveraging cellular infrastructure to support large-scale sensing without additional hardware deployment. This paper presents the first comprehensive survey dedicated to MIMO OFDM-enabled ISAC for low-altitude non-cooperative UAV surveillance, where the targeted UAVs do not intentionally assist the monitoring system through dedicated signaling or prior coordinate sharing. We first analyze the unique propagation characteristics of low-altitude UAV sensing, including severe clutter, rapid channel variations, and mixed near/far-field effects, and discuss corresponding waveform design principles. We then systematically review existing MIMO OFDM-enabled UAV surveillance techniques along four key dimensions: ISAC system modeling and network optimization, UAV detection and tracking algorithms under single and networked base station (BS) architectures, UAV identification techniques based on micro-Doppler and learning-based approaches, and experimental validations and practical field trials. Subsequently, we summarize open challenges such as sensing under severe clutter and multipath, data scarcity for identification, cooperative multi-BS fusion, and real-world deployment constraints. Finally, we outline promising future research directions toward 5G-Advanced (5G-A) and 6G-enabled low-altitude surveillance systems.
\end{abstract}


\section{Introduction}
\IEEEPARstart{R}{recently}, the growing utilization of unmanned aerial vehicles (UAVs) has greatly improved operational efficiency and flexibility across various fields, such as logistics, environmental monitoring, and aerial mapping \cite{wandelt2023aerial, ruckin2023informative}. However, their proliferation has also raised safety and security concerns \cite{he2024detection}, as UAVs can be exploited for malicious purposes through unauthorized flights, highlighting the need for early threat identification as a prerequisite for effective countermeasures. To this end, various dedicated sensing systems have been developed for UAV surveillance, including radar-based systems, vision-based systems, and multi-modal sensing systems \cite{dong2025securing}. In spite of their effectiveness under certain conditions, these systems are often limited by environmental variations and high deployment costs, particularly for large-scale implementation.

Beyond these dedicated sensing systems, recent advances in wireless communications have opened up an alternative route for UAV surveillance. In particular, with the increasing adoption of wideband orthogonal frequency division multiplexing (OFDM) waveforms and multiple-input multiple-output (MIMO) techniques in 5G and beyond, communication signals can provide high resolution in both time and spatial domains \cite{liu2022survey}. This capability enables high-accuracy sensing using communication waveforms, thereby facilitating the paradigm of integrated sensing and communication (ISAC) \cite{buzzi2019using}. In general, the advantages of adopting ISAC for UAV surveillance are twofold. First, ISAC is less sensitive to illumination conditions and can operate more reliably in diverse environments compared with vision-based surveillance systems \cite{huang2025cooperative}. Second, by reusing the existing cellular infrastructure, ISAC enables wide-area and persistent surveillance with lower incremental deployment cost, making it a promising solution for non-cooperative UAV monitoring.

In fact, ISAC has been widely recognized as a key enabling technology in the future 5G-Advanced (5G-A) and 6G mobile network, with potential applications in areas such as smart home, intelligent transportation, and vehicle to everything (V2X) \cite{cui2021integrating, bayesteh2022integrated}. However, ISAC for UAV surveillance exhibits several distinctive characteristics, which can be summarized as follows:
\begin{itemize}
    \item \textbf{Small size and high mobility:}
    Unlike other conventional sensing targets, UAVs are typically small in size and highly mobile in three-dimensional low-altitude airspace. The small physical size of UAVs often leads to a relatively small radar cross section (RCS), which results in weak target echoes, reduced sensing signal-to-noise ratio (SNR). Meanwhile, the high mobility causes rapid channel variations, which complicates reliable detection, localization, and continuous tracking, especially in cluttered low-altitude environments.
    \item \textbf{Wide-area and persistent monitoring requirements:}
    Many existing ISAC scenarios mainly focus on local environment perception or link-level sensing. In contrast, UAV surveillance requires continuous monitoring over large geographic areas for long durations, often relying on infrastructure reuse and cooperation among multiple base stations (BSs) to ensure sufficient coverage and tracking continuity. Moreover, since UAVs may appear anywhere in the airspace and can fall into either the near-field or far-field region of different BSs, the sensing network must be capable of supporting both sensing regimes while providing ubiquitous coverage.
    \item \textbf{A complete surveillance chain beyond sensing only:}
    UAV surveillance is not limited to estimating target range or velocity. Instead, it usually involves a complete functional chain, including detection, localization, tracking, identification, and even integration with airspace management systems. This makes UAV surveillance a more system-level problem than other conventional ISAC applications.
\end{itemize}

Consequently, ISAC systems for UAV surveillance require robust designs at both the physical and network layers. First, the ISAC waveform used for UAV surveillance should adapt to rapid channel variations and be capable of maintaining reliable sensing performance with weak target echoes. As the dominant waveform in current cellular networks, OFDM achieves remarkable performance in both communication and sensing, with favorable delay and Doppler resolution enabled by its thumbtack-shaped ambiguity function \cite{dai2026tutorial}. In recent years, several emerging ISAC waveforms, including delay-Doppler alignment modulation (DDAM) \cite{xiao2023integrated, xiao2024integrated-ddam}, orthogonal time frequency space (OTFS) \cite{yuan2021integrated}, orthogonal chirp division multiplexing (OCDM) \cite{ouyang2016orthogonal}, and affine frequency division multiplexing (AFDM) \cite{bemani2023affine}, have been proposed, with each offering distinct advantages in different aspects especially in high mobility scenarios. However, the complex signal processing and high computational cost associated with these waveforms make them less suitable for UAV surveillance scenarios with stringent real-time requirements. Second, the sensing network should support distributed target monitoring, cooperative beam scheduling, and seamless sensing handover across different BSs. Numerical studies have demonstrated the performance of ISAC in a networked sensing framework with cooperative beamforming \cite{mao2024communication, ma2022cooperative, zhang2024cooperative}. However, the sensing capability with more complicated wireless channel, such as mixed near-field and far-field channel, requires further exploration.

In this context, MIMO OFDM-enabled cellular ISAC is particularly relevant and practically important for low-altitude non-cooperative UAV surveillance. On the one hand, MIMO provides high spatial resolution and flexible beamforming capability, which are essential for angular localization, target separation, and interference suppression. On the other hand, in addition to its wideband characteristic that substantially improves timing resolution, OFDM offers structured time-frequency resources that facilitate the extraction of delay and Doppler information from communication waveforms. More importantly, the cellular infrastructure itself enables wide-area and persistent surveillance through infrastructure reuse, low incremental deployment cost, and cooperation among multiple base stations. Therefore, the importance of MIMO OFDM-enabled cellular ISAC lies not only in its theoretical exclusivity, but in its strong alignment with the practical requirements and deployment realities of low-altitude non-cooperative UAV surveillance. Therefore, focusing on MIMO OFDM-enabled ISAC for low-altitude non-cooperative UAV surveillance, this work systematically reviews the state of the art in this field. Major abbreviations used in this article are summarized in Table \ref{tab:1}.

\begin{table*}[hbt]
\centering
\caption{Major acronyms used in this paper.}
\label{tab:1}
\begin{tabular}{cccc}
\toprule[1pt]
%\hline\noalign{\smallskip}
\textbf{\makecell{Acronym}} & \textbf{\makecell{Definition}} & \textbf{\makecell{Acronym}} & \textbf{\makecell{Definition}} \\
\noalign{\smallskip}\hline\noalign{\smallskip}
UAV & Unmanned aerial vehicle & mmWave & Millimeter wave \\
MIMO & Multiple-input multiple-output & ISAC & Integrated sensing and communication \\
5G-A & 5G-Advanced & OFDM & Orthogonal frequency division multiplexing \\
RIS & Reconfigurable intelligent surface & ML & Machine learning \\
RCS & Radar cross-section & BS & Base station \\
RF & Radio frequency & ACF & Auto-correlation function \\
UE & User equipment & SSB & Synchronization signal block \\
NR & New radio & RS & Reference signal \\
DFT & Discrete Fourier transform & FFT & Fast Fourier transform \\
PRS & Positioning reference signal & PDSCH & Physical downlink shared channel \\
3GPP & 3rd generation partnership project & DMRS & Demodulation reference signal \\
IFFT & Inverse fast Fourier transform & CRLB & Cram\'er-Rao lower bound \\
QAM & Quadrature amplitude modulation & PSK & Phase shift keying \\
SCA & Successive convex approximation & SNR & Signal-to-noise ratio \\
CRB & Cram\'er-Rao bound & DoA & Direction-of-arrival \\
MLE  & Maximum likelihood estimation & MSE & Mean squared error \\
GLRT & Generalized likelihood ratio test & SCNR & Signal-to-clutter-plus-noise ratio \\
OMP & Orthogonal matching pursuit & DFBS & Dual-functional base station \\
FOV & Field of view & OTA & Over-the-air \\
PD & Probability of detection & EKF	& Extended Kalman filter \\
GBS & Ground base station & SINR & Signal-to-interference-plus-noise ratio \\
QoS & Quality of service & MCBPA & Multi-BS collaborative beam power allocation \\
AO & Alternating optimization & FP & Fractional programming  \\
LoS & Line of sight & SDR & Semi definite relaxation \\
ISAR & Inverse synthetic aperture radar & PSO & Particle swarm optimization \\
MCRB & Misspecified Cram\'er-Rao bound & ESPRIT & Estimating signal parameters via rotational variation techniques \\
IMMUKF & Interactive multiple-model unscented Kalman filter & UKF & Unscented Kalman filter \\
VSC & Virtual sensing cell & MUSIC & Multiple signal classification \\
LAE & Low-altitude economy & AoA & Angle of arrival \\
CS & Compressed sensing & RMSE & Root mean square error \\
AP & Access point & CSI & Channel state information \\
CNN & Convolutional neural network & TDD & Time-division duplex \\
rmD-NSP & Rotor micro-Doppler null space pursuit & STFT & Short-time Fourier transform \\
RTK & Real-time kinematic & GPS & Global positioning system \\
IoT & Internet of Things & SLAM & Simultaneous localization and mapping \\
CDF & Cumulative distribution function & MTI & Moving target indicator \\
GAN & Generative adversarial networks & LLM & Large language model \\
\bottomrule[1pt]
\end{tabular}
\end{table*}

\subsection{Existing Surveys and Our Contributions}
Recently, several surveys have provided overviews of ISAC-based sensing technologies from different perspectives.

\textit{1) General ISAC Surveys:} Aldirmaz-Colak \textit{et al.} provided a comprehensive survey of ISAC for 6G networks from both sensing-centric and communication-centric perspectives \cite{aldirmaz2025comprehensive}. Their study reviews key ISAC enablers such as waveform design, edge computing, and vehicular communications, and emphasizes security considerations as well as emerging directions including reconfigurable intelligent surface (RIS), non-orthogonal multiple access (NOMA), and UAVs. It also highlights the rapidly growing role of artificial intelligence (AI) and machine learning (ML) in next-generation ISAC systems. However, this survey focuses on general ISAC architectures and does not analyze sensing techniques tailored for non-cooperative UAV surveillance. Similarly, Wei \textit{et al.} provided a comprehensive overview of ISAC signal design, processing, and optimization \cite{wei2023integrated}. However, UAV sensing and detection are only mentioned briefly as one of the potential future application scenarios, without detailed methodological analysis.

\textit{2) ISAC for UAV \& Low-Altitude Surveys:} More recently, several studies have begun exploring ISAC in low-altitude or UAV-related scenarios. Ma \textit{et al.} presented a comprehensive survey on ISAC network design for the low-altitude economy, discussing key challenges such as detecting UAVs with small RCS, interference management, sensing and communication coverage mismatch, and air-ground coordination \cite{dingyou2025integrated}. While their study highlights non-cooperative UAV intrusion detection and micro-Doppler-based recognition, the analysis remains at a system-level perspective without delving into signal processing algorithms for UAV surveillance within a unified ISAC system. Mu \textit{et al.} reviewed ISAC-enabled UAV systems and identified non-cooperative UAV detection as an emerging direction in low-altitude sensing \cite{mu2023uav}. They show that ISAC BSs can detect unauthorized UAVs from downlink echoes or UAV control-link signals, and emphasize AI-assisted fusion (e.g., federated and transfer learning). Song \textit{et al.} offered an overview of cellular ISAC for low-altitude UAVs, highlighting UAVs' dual roles as sensing targets and aerial anchors, but the discussion remains largely conceptual without detailed algorithmic analysis for non-cooperative UAV surveillance \cite{song2025overview}. In contrast, Ahmed \textit{et al.} focused on UAV-assisted ISAC \cite{ahmed2025advancements}, where UAVs act as active sensing nodes to enhance situational awareness in complex environments such as disaster zones or urban canyons, rather than serving as non-cooperative targets.

\textit{3) Anti-UAV Surveys:} Parallel to ISAC-oriented works, several surveys have summarized the development of anti-UAV systems from a broader multi-sensor perspective. Dong \textit{et al.} reviewed UAV detection, classification, and tracking methods using electro-optical, infrared, radar, radio frequency (RF), and acoustic modalities, with a strong emphasis on multi-sensor fusion and deep learning pipelines rather than ISAC-based sensing \cite{dong2025securing}. Khawaja \textit{et al.} systematically analyzed UAV detection, classification, and tracking using radar, communication signals, RF analyzers, and non-RF sensors, focusing on traditional RF sensing rather than ISAC-enabled frameworks \cite{khawaja2025survey}.

\begin{figure*}[hbt]
\centering
\includegraphics[scale=0.75]{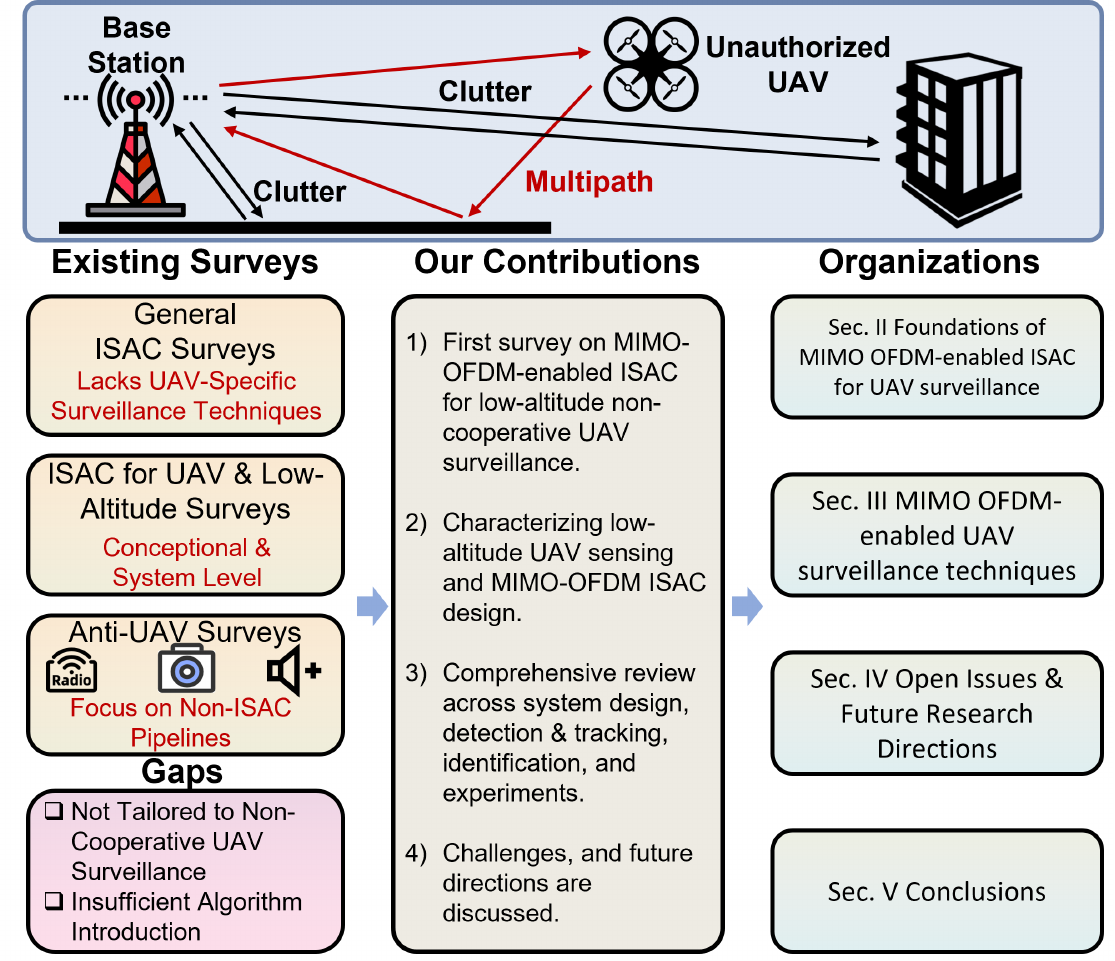}
\caption{Overview of this paper.}
\label{fig:1}
\end{figure*}

In summary, while existing surveys have provided valuable insights into ISAC theory, multi-sensor anti-UAV systems, and UAV-enabled sensing platforms, none has specifically addressed MIMO OFDM-enabled ISAC architectures for low-altitude non-cooperative UAV detection, tracking, and identification. The comparison between existing surveys and this paper is provided in Table \ref{tab:2}. The overview of this paper is shown in Fig. \ref{fig:1}.

\begin{table*}[hbt]
\renewcommand{\arraystretch}{1.5}
\centering
\caption{Comparison of previous works and this survey.}
\label{tab:2}
\begin{threeparttable}
\begin{tabular}{c|c|c|c|c|c}
\toprule[1pt]
\textbf{\makecell{References}} &
\textbf{\makecell{Waveform Specificity}} &
\textbf{\makecell{BS Architecture}} &
\textbf{\makecell{Processing Scope}} &
\textbf{\makecell{Experimental Validation \\ Coverage}} &
\textbf{\makecell{UAV-specific Focus}}\\
\cline{1-6}
\cite{dong2025securing} & -- & -- & \makecell{Detection, Tracking, \\ Identification, \& Benchmarking} & \ding{51} & \ding{51}  \\
\cline{1-6}
\cite{aldirmaz2025comprehensive} & Multiple & Single \& Multi-BS & \makecell{Signal Processing, \\ \& Optimization} & \ding{55} & \ding{55}  \\
\cline{1-6}
\cite{wei2023integrated} & Multiple & -- & \makecell{Signal Processing, \\ \& Optimization} & \ding{55} & \ding{55} \\
\cline{1-6}
\cite{dingyou2025integrated}& -- & Multi-BS & System-level Analysis & \ding{55} & \ding{51} \\
\cline{1-6}
\cite{mu2023uav}& -- & -- & System-level Analysis & \ding{55} & \ding{51}  \\
\cline{1-6}
\cite{song2025overview} & -- & Multi-BS & System-level Analysis & \ding{55} & \ding{51}  \\
\cline{1-6}
\cite{ahmed2025advancements} & Multiple & -- & \makecell{Signal Processing, \\ \& Optimization} & \ding{55} & \ding{51} \\
\cline{1-6}
\cite{khawaja2025survey} & Multiple & -- & \makecell{Detection, Tracking, \\ \& Identification} & \ding{51} & \ding{51}  \\
\cline{1-6}
\text{Our work} & OFDM & Single \& Multi-BS & \makecell{Signal Processing, Detection, Tracking, \\ \& Identification} & \ding{51} & \ding{51}  \\
\bottomrule[1pt]
\end{tabular}
\end{threeparttable}
\end{table*}

Therefore, the main contributions of this article are as follows:
\begin{itemize}
  \item [1)]
  To the best of our knowledge, this is the first survey that systematically reviews research on MIMO OFDM-enabled ISAC for low-altitude, non-cooperative UAV surveillance.
  \item [2)]
  This paper analyzes the unique signal characteristics and propagation properties of low-altitude UAV sensing, and summarizes corresponding waveform design and receiver processing considerations for MIMO OFDM-based ISAC systems.
  \item [3)]
  This paper presents a comprehensive and in-depth survey of MIMO OFDM-enabled UAV surveillance techniques, organized along four key dimensions: system design and optimization, detection and tracking, identification, and corresponding experimental tests and analyses.
  \item [4)]
  This paper discusses the limitations and challenges of existing methods, and concludes by outlining promising directions for future research.
\end{itemize}

\subsection{Organization of this Article}
The rest of this article is structured as follows. Section \ref{sec2} presents the foundations of MIMO OFDM-enabled ISAC for low-altitude UAV surveillance. In Sections \ref{sec3}, MIMO OFDM-enabled UAV surveillance techniques are introduced. In Section \ref{sec4}, open issues and future research trends are presented. Finally, Section \ref{sec5} concludes this article.

\section{Foundations of MIMO OFDM-Enabled ISAC for UAV Surveillance}
\label{sec2}
In this section, we discuss the foundations of MIMO OFDM-enabled ISAC with a particular focus on UAV surveillance. We first identify the key factors that render UAV surveillance fundamentally different from other widely studied ISAC applications. We then review state-of-the-art waveform design approaches that apply to UAV monitoring. Finally, we categorize representative signal processing algorithms according to different propagation models.

\subsection{Received Signal Characteristics for UAV ISAC}
For UAV-oriented ISAC, it is crucial to first examine the characteristics of the received signal. Compared with conventional ISAC scenarios, the signal properties are significantly altered by the low-altitude operation, which in turn leads to specific design requirements. To this end, we summarize these requirements to guide the subsequent MIMO OFDM waveform design and the development of receiver signal processing algorithms. An overview of the propagation properties for UAV ISAC is illustrated in Fig. \ref{fig:2}.

\subsubsection{Low-Altitude Clutter and Multipath}
The low-altitude ISAC environment enables rich signal reflections and scattering from the ground and nearby objects such as buildings, trees, and vehicles, as shown in Fig. \ref{fig:2a}. Consequently, in addition to the desired echo from the target UAV, the receiver also observes low-altitude clutter and multipath components that may carry a substantial portion of the total received signal energy \cite{wu2025toward}. On the one hand, the clutter introduces strong interference that elevates the sidelobe level of the auto-correlation function (ACF), thereby severely degrading detection performance. On the other hand, the multipath components induce structured and time-varying interference, which often leads to non-Gaussian statistics of the received signal. This, however, induces statistical uncertainties and makes performance evaluation and false-alarm control more challenging.

\subsubsection{Rapid Time-Varying Channel}
Unlike conventional ISAC settings where a quasi-static channel model can be assumed over a short observation interval, the sensing channel for UAV detection varies rapidly and exhibits pronounced frequency selectivity \cite{liu2022characterizations}. This is mainly attributed to two factors. First, the high mobility of the UAV induces significant Doppler shifts and Doppler spread. This shortens the channel coherence time and leads to rapid variations in the phase and amplitude of the received signal, thereby reducing the stability and separability of target-related characteristics in the delay-Doppler domain. In addition, the Doppler spread smears the received energy across Doppler bins, which degrades the peak-to-sidelobe ratio (PSLR) and increases detection ambiguity. Secondly, in dense urban environments, frequent blockages cause propagation paths to appear and disappear over time, making the received signal intermittent and leading to abrupt changes in the channel response, as shown in Fig. \ref{fig:2b}. As a result, the observations become less consistent over the sensing interval, and the signal statistics are more difficult to characterize.

\subsubsection{Near-field and Far-field Coexistence}
\begin{figure*}[htb]
    \centering
    \subfloat[Low-altitude scattering\label{fig:2a}]{
        \includegraphics[width=0.29\linewidth]{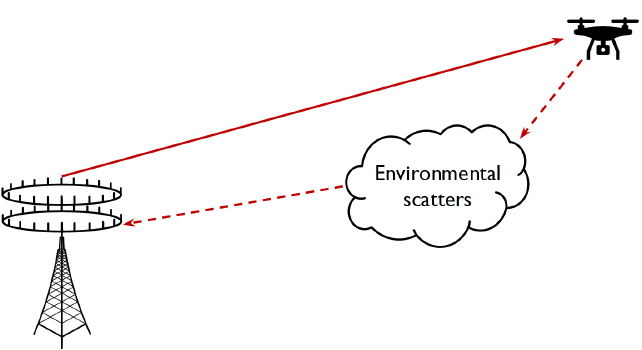}
    }\hfill
    \subfloat[Signal intermittent due to blockage\label{fig:2b}]{
        \includegraphics[width=0.29\linewidth]{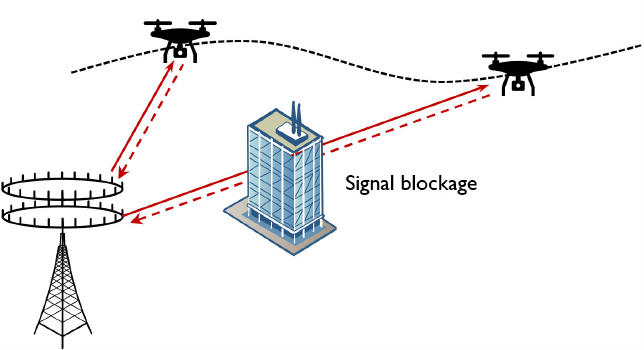}
    }\hfill
    \subfloat[Near-field and far-field coexistence\label{fig:2c}]{
        \includegraphics[width=0.29\linewidth]{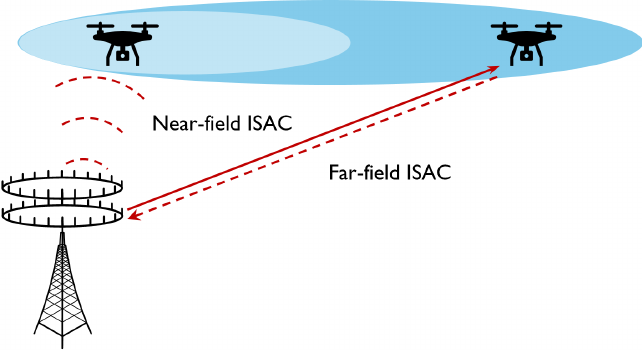}
    }
    \caption{Illustration of the key characteristics of UAV-ISAC.}
    \label{fig:2}
\end{figure*}

The Rayleigh distance separates the electromagnetic waves radiated from antennas into two regions, commonly referred to as the near-field and far-field regions \cite{liu2023near}. In conventional cellular deployments, far-field considerations with planar waves are often adopted since the typical link distances are much larger than the corresponding Rayleigh distance. However, in mmWave MIMO ISAC systems, the joint effect of large-aperture arrays and short wavelengths leads to an increase in the Rayleigh distance, so that the sensing targets are likely to fall into the near-field region, where the spherical wavefronts may bring challenges to the sensing mission. Consider a typical 5G mmWave cellular deployment, where a BS is equipped with $N=128$ antenna elements and operates at a carrier frequency $f_c=28$ GHz. The corresponding Rayleigh distance is given by $2D^2/\lambda\approx86.4$ m, where $D=(N-1)/\lambda$ denotes the array aperture. For the surveillance of ground vehicles along a street, the sensing operation often remains in a single propagation regime over the region of interest, which can be determined by the BS placement and the street geometry. In contrast, UAV surveillance exhibits a less constrained spatial footprint due to its flexible mobility. Consequently, the BS receiver may operate in a mixed near-far-field regime as UAVs can easily move across a wide range of distances without limitations, as shown in Fig. \ref{fig:2c}. This necessitates a unified signal processing capability for the sensing receiver that remains effective under a mixed-field propagation condition.

Against the above discussion, it can be concluded that different from other sensing scenarios, MIMO OFDM-enabled ISAC for UAV surveillance necessitates unique physical-layer design principles, which are outlined as follows:
\begin{itemize}
    \item \textbf{Clutter and Multipath Robustness:}  The waveform used for UAV surveillance should maintain a low correlation sidelobe floor to preserve a clear detection under strong low-altitude clutter. In addition, the frame structure should provide adequate time-frequency sensing opportunities for multi-symbol accumulation to mitigate multipath effects.
    \item \textbf{Time-Frequency Selectivity Awareness:} On the one hand, the waveform design should be time-selectivity-aware and incorporate periodic redundancy to cope with intermittent receptions caused by signal blockage. On the other hand, it should be frequency-selectivity-aware and provide sufficient frequency-domain resources for sensing to maintain stable measurement and reliable modeling. Moreover, the receiver design should account for time-varying Doppler shifts and Doppler spread, which can cause potential harm to the orthogonality of OFDM structure, thereby inducing inter-carrier interference (ICI).
    \item \textbf{Mixed-Field Compatibility:} The sensing receiver should not be restricted to a single propagation regime, i.e., near-field or far-field. Since the UAV may appear either in the close proximity of the BS or at much longer ranges, the processing pipeline should remain effective in a mixed field setting and support both spherical wavefront for near-field ISAC and planar wavefront for far-field ISAC.
\end{itemize}

\subsection{Signaling Design for MIMO OFDM ISAC}
In general, waveform design strategies for dual-functional ISAC systems can be categorized into three main categories. First, sensing-centric designs aim to maximize the sensing performance by embedding communication data into traditional radar waveforms \cite{hassanien2015dual, huang2020majorcom}. Secondly, communication-centric designs reuse existing OFDM-based communication waveforms to ensure full compatibility with current cellular deployments \cite{du2024reshaping, wu2022integrating}. Thirdly, joint sensing and communication designs simultaneously account for communication requirements and sensing performance metrics, and optimize the transmit signaling accordingly, typically via dedicated optimization frameworks \cite{liu2018toward}. Although other signaling methods may achieve better sensing performance, for MIMO OFDM-based ISAC, the communication-centric designs are particularly attractive, as they can be implemented on the existing cellular infrastructure without requiring large-scale hardware replacement. To this end, the remainder of this section reviews signaling designs that reuse the communication waveform for sensing, including both deterministic pilot signals and random data symbols. The summary of signaling design for MIMO OFDM ISAC is shown in Table \ref{tab:waveform}.

\subsubsection{Sensing with deterministic pilot signals}
In cellular communication systems, pilot signals are sparsely embedded in the time-frequency resources, and their structures are known to both the BS and the user equipment (UE) to facilitate channel estimation. These signals are typically constructed from sequences with good auto-correlation properties, such as Zadoff-Chu (ZC) sequences and m-sequences, to facilitate channel estimation. Although sensing relies on a different set of performance metrics compared with channel estimation, existing studies have demonstrated that these pilot signals can still be effectively utilized for sensing purposes \cite{wei2024multiple, cui2022integrated}.

For downlink sensing in monostatic scenarios, the synchronization signal block (SSB), which is originally used for initial access and synchronization, has been investigated as a sensing waveform \cite{chen2024isac, du2025simultaneous, abratkiewicz2023ssb}. Specifically, Chen \textit{et al.} investigated an ISAC-enabled beam alignment framework for THz networks that reuses standard 5G new radio (NR) SSBs and reference signals (RSs) for sensing \cite{chen2024isac}. The authors propose a joint SSB-RS scheme, where periodic SSBs are exploited for blockage detection and coarse sensing, while comb-type RSs embedded in data transmission are used for fine range-velocity estimation and user tracking. They analytically characterize the sensing performance in terms of range and velocity resolution and unambiguous intervals as functions of the sensing time-frequency pattern and derive design guidelines for configuring the resources to minimize beam misalignment. Numerical results show that the proposed design significantly improves beam alignment robustness and coverage probability compared with conventional communication-only baselines. Besides, Du \textit{et al.} proposed an active mmWave ISAC framework that utilizes SSB for simultaneous localization and mapping (SLAM) \cite{du2025simultaneous}. For each beam direction, the correlation is performed between the received echo and the local sequence, thereby obtaining the coarse delay-based range estimates. The coarse estimates are further refined using the OFDM channel response, power-delay profile, and a binary search procedure. By steering beams over a discrete Fourier transform (DFT) codebook and extracting dominant delay peaks, the system builds radio point clouds across multiple viewpoints, which are then processed via point-cloud registration and pose-graph optimization to jointly estimate the terminal trajectory and reconstruct a detailed radio map.

In addition, Abratkiewicz \textit{et al.} proposed an SSB-based signal processing framework for 5G passive radar, targeting the initial access procedure \cite{abratkiewicz2023ssb}. The authors first decode and reconstruct the complete SSB at the passive radar node and use the synthesized SSB as a matched-filter reference to form range-time maps, followed by phase alignment, clutter cancellation, and slow-time fast Fourier transform (FFT) to obtain range-Doppler maps. Furthermore, the ambiguity function and correlation metrics are analytically modelled to show that using the full reconstructed SSB yields superior sidelobe and resolution properties compared with using only partial SSB components. In this regard, the authors further propose a simple two-snapshot method to resolve Doppler ambiguity in single-target cases. Simulations and real-world experiments with vehicular targets confirm that the SSB-based processing can reliably detect moving objects even in the absence of downlink content, where classical cross-ambiguity function (CAF)-based passive coherent location (PCL) processing fails, thereby significantly extending the usability of 5G-based passive radar in low-traffic short-range scenarios.

In addition to SSB-based sensing methods, the use of positioning reference signal (PRS) in monostatic ISAC is also attractive, as it provides a larger effective bandwidth in the frequency domain, thereby enabling a higher ranging accuracy \cite{khosroshahi2024leveraging, wei20225g}. In particular, Khosroshahi \textit{et al.} investigated an ISAC framework that leverages PRS for sensing and the physical downlink shared channel (PDSCH) for communication, with the goal of enabling passive target localization without modifying 3rd generation partnership project (3GPP)-compliant waveforms \cite{khosroshahi2024leveraging}. It can be shown that the comb-pattern structure of PRS leads to range ambiguity and ghost targets. Therefore, two demodulation reference signal (DMRS)-assisted algorithms for different PRS comb sizes are proposed, which jointly exploit PRS and PDSCH-DMRS to suppress ghost peaks while preserving range resolution and maximum detectable range. Followed by this, an OFDM resource grid carrying both PRS and PDSCH is formulated as a resource-allocation problem to identify Pareto-optimal tradeoffs between communication throughput and sensing accuracy under ghost-free operation. Simulation results demonstrate that the proposed design effectively removes ghost targets, and the optimized time–frequency allocation achieves a desirable balance between sensing performance and communication rate in 5G-compliant ISAC systems. Wei \textit{et al.} investigated a 5G PRS-based sensing reference signal approach for joint sensing and communication by using 5G PRS \cite{wei20225g}. In this paper, a PRS-based OFDM radar scheme is proposed for range and velocity estimation, including ambiguity-function analysis and fractional inverse fast Fourier transform (IFFT)/FFT techniques to enhance estimation accuracy. Meanwhile, they derive the Cram\'er-Rao lower bound (CRLB)s for PRS-based range and velocity sensing as well as for PRS-based positioning and further introduce a multi-frame velocity estimation method to improve accuracy while controlling pilot overhead. Simulation results not only demonstrate that PRS offers superior sensing performance over other pilots, but provide design insights for future 5G-A and 6G frame structures with dedicated sensing reference signals.

\subsubsection{Sensing with random data symbols}
Although pilot-based sensing achieves considerable performance in the aforementioned works, it is generally inadequate for high-mobility UAV surveillance due to the sparse time-frequency structure of these pilot signals. To obtain additional sensing resources for multi-symbol accumulation to maintain enhanced Doppler resolution and detection reliability, several recent studies have investigated exploiting random data payloads for sensing \cite{lu2024random, liu2025uncovering}. The idea of utilizing random data payloads for sensing was initially proposed by Lu \textit{et al.} \cite{lu2024random}, where a new sensing metric, the ergodic linear minimum mean square error (ELMMSE), is proposed to measure the estimation error averaged over random signal realizations. In addition, two dedicated precoding schemes are developed for sensing-only scenarios: a data-dependent precoder that adapts to each instantaneous data realization, and a lower-complexity data-independent precoder optimized via stochastic gradient projection. They further extend these designs to joint sensing and communication operation by incorporating a communication rate constraint and developing a penalty-based alternating optimization (AO) framework. Numerical results show that the proposed random signal aware precoders substantially outperform conventional ISAC designs that treat the signal covariance as deterministic, thereby demonstrating that random ISAC signals merit dedicated precoding strategies.

To further characterize the sensing performance enabled by random communication data, the authors in \cite{liu2025uncovering} investigated the fundamental impact of modulation and pulse shaping. They derive a closed-form expression for the expected squared ACF of random ISAC waveforms under arbitrary orthonormal modulation bases and Nyquist pulse shaping, revealing the so-called “iceberg in the sea” structure, where a pulse-dependent squared mean (“iceberg”) and a data-induced variance term (“sea level”) jointly determine the ranging sidelobes. It is shown that, for sub-Gaussian constellations such as quadrature amplitude modulation (QAM) and phase shift keying (PSK), OFDM is the unique modulation basis that minimizes ranging sidelobe levels for all delays. Building on this insight, an \textit{iceberg shaping} approach is proposed to design Nyquist pulses that further suppress sidelobes within a desired delay region, particularly when combined with coherent integration. Numerical results demonstrate that the proposed pulse shaping can substantially reduce sidelobe levels and improve ranging accuracy compared with conventional root-raised cosine (RRC) pulse shaping.

Nevertheless, sensing with random data symbols generally exhibits inferior performance compared with deterministic sensing waveforms, owing to the lack of a known sequence and weaker auto-correlation properties. This has motivated a line of research that investigates sensing performance when jointly exploiting random information signals and dedicated deterministic sensing waveforms \cite{xu2025exploiting, song2025crb, xie2025bistatic}. Xu \textit{et al.} investigated an ISAC framework that jointly exploits both deterministic pilots and random data payload symbols for sensing in a monostatic MIMO system \cite{xu2025exploiting}. The authors adopt the ELMMSE proposed in \cite{lu2024random} as the sensing performance metric. By leveraging random matrix theory, derive semi-closed asymptotic expressions that are independent of specific data realizations, thereby enabling tractable analysis and precoder design. Based on these expressions, they formulate a precoding optimization problem to minimize the sensing error under transmit-power and communication-rate constraints and develop an successive convex approximation (SCA)-based algorithm together with a high-SNR closed-form approximation that yields a convex problem with a globally optimal solution. Numerical results show that properly reusing data payload symbols can reduce the sensing error by up to 5.6 dB compared with pilot-only sensing, while incurring almost no loss in communication performance. The Cram\'er-Rao bound (CRB)-rate tradeoff was investigated in \cite{song2025crb}, where a multi-antenna BS serves a communication user while a separate multi-antenna sensing receiver estimates the target direction-of-arrival (DoA) using either Gaussian information signals alone or a superposition of Gaussian information and deterministic sensing signals. The authors first derive closed-form CRBs for DoA estimation under the practical assumption that the sensing receiver only knows the covariance of the Gaussian information signals, and design corresponding maximum-likelihood estimators (MLEs) whose mean squared error (MSE) approaches these bounds at high sensing SNR. Based on the derived CRBs, the authors formulate transmit beamforming optimization problems to minimize the sensing CRB under signal-to-interference-plus-noise ratio (SINR) and power constraints in both sensing-only and joint S\&C modes, and obtain a convex optimal solution for the Gaussian-only case as well as an SCA-based iterative algorithm when deterministic sensing signals are also transmitted. Numerical results show that superimposing deterministic sensing signals is crucial for improving sensing accuracy, and the proposed beamforming designs yield a superior ISAC performance boundary compared with several benchmark schemes.

In addition, Xie \textit{et al.} investigated bistatic target detection in ISAC systems that exploit both deterministic pilots and unknown random data payloads in a hybrid transmit signal \cite{xie2025bistatic}. The authors propose a generalized likelihood ratio test (GLRT)-based detector tailored to this hybrid model, where the target echo induces coupled changes in both the mean and covariance of the received signal, while the sensing receiver only obtains the pilots and the statistical knowledge of the data symbols. They derive asymptotic closed-form expressions for the false-alarm and detection probabilities in the large-sample regime, which reveal a fundamental tradeoff: deterministic pilots and random data both enhance detection reliability, but the data component also introduces statistical uncertainty that can degrade performance. Simulation results validate the accuracy of the asymptotic analysis and demonstrate that the proposed detector significantly outperforms pilot-only and data-only baselines, especially when the number of available samples is moderate, thereby underscoring the importance of dedicated detectors that fully exploit data payloads for sensing.

To sum up, ISAC for UAV surveillance can be practically realized using either deterministic reference signals only or data-assisted transmission signals, with different tradeoffs in sensing capability, mobility support, signaling overhead, and standard compliance. The comparison between the aforementioned sensing signals is summarized in Table \ref{tab:compare}.

\begin{table*}[hbt]
\renewcommand{\arraystretch}{1.3}
\centering
\caption{Summary of Signaling Design for MIMO OFDM ISAC.}
\label{tab:waveform}
\begin{threeparttable}

\begin{tabular}{
    >{\centering\arraybackslash}m{2.5cm}|
    >{\centering\arraybackslash}m{2.5cm}|
    >{\centering\arraybackslash}m{10cm}
}
\toprule[1pt]
\textbf{Reference} & \textbf{Sensing Signal} & \textbf{Key Contributions} \\
\midrule

{\cite{du2025simultaneous, abratkiewicz2023ssb}} & SSB &
\begin{itemize}[leftmargin=*]
  \item An active mmWave ISAC-SLAM platform is provided with real-world validation.
  \item A passive radar processing framework with 5G-SSB signals is developed.
\end{itemize}
\\ \cline{1-3}

{\cite{wei20225g}} & PRS &
\begin{itemize}[leftmargin=*]
  \item CRLBs for PRS-based ranging and velocity sensing are derived.
\end{itemize}
\\ \cline{1-3}

{\cite{khosroshahi2024leveraging}} & PRS + DMRS &
\begin{itemize}[leftmargin=*]
  \item A joint processing scheme is developed to suppress ghost peaks while preserving range resolution and maximum detectable range.
\end{itemize}
\\ \cline{1-3}

{\cite{chen2024isac}} & SSB + PRS &
\begin{itemize}[leftmargin=*]
  \item A two-stage sensing design is proposed for THz ISAC beam alignment.
\end{itemize}
\\ \cline{1-3}

{\cite{lu2024random, liu2025uncovering}} & Random signal &
\begin{itemize}[leftmargin=*]
  \item A random-data-aware ISAC precoding framework is proposed, with dedicated data-dependent and data-independent precoders extended to joint sensing-communication design.
  \item A closed-form sidelobe analysis for random ISAC waveforms is derived, and an iceberg-shaping pulse design for sidelobe suppression is established.
\end{itemize}
\\ \cline{1-3}

{\cite{xu2025exploiting, song2025crb, xie2025bistatic}} & Random signal + pilot signal &
\begin{itemize}[leftmargin=*]
  \item A pilot-data joint sensing precoding design is developed using asymptotic ELMMSE analysis for tractable optimization.
  \item Closed-form DoA CRB analysis and CRB-driven beamforming designs are developed for hybrid deterministic/random ISAC signaling.
  \item A GLRT-based hybrid pilot-data detector is proposed, with asymptotic detection analysis for bistatic ISAC target detection.
\end{itemize}
\\

\bottomrule[1pt]
\end{tabular}

\end{threeparttable}
\end{table*}

\begin{table*}[hbt]
\renewcommand{\arraystretch}{1.3}
\centering
\caption{Comparison between different sensing signals.}
\label{tab:compare}
\begin{threeparttable}

\begin{tabular}{
    >{\centering\arraybackslash}m{2cm}|
    >{\centering\arraybackslash}m{4cm}|
    >{\centering\arraybackslash}m{2cm}|
    >{\centering\arraybackslash}m{2cm}|
    >{\centering\arraybackslash}m{2cm}|
    >{\centering\arraybackslash}m{4cm}
}
\toprule[1pt]
\textbf{Sensing Signal} & \textbf{Potential Sensing Accuracy} & \textbf{Mobility Robustness} & \textbf{Signaling Overhead} & \textbf{Standard Compatibility} & \textbf{Suitability for UAV Sensing}\\
\midrule

SSB & \begin{itemize}[leftmargin=*]
    \item Primarily suitable for coarse sensing and initial target discovery
    \item Limited sensing accuracy due to sparse transmission and restricted resource allocation
\end{itemize}
&
Medium & Very low & Very high & \begin{itemize}[leftmargin=*]
    \item Initial detection
    \item Coarse localization
    \item Beam acquisition
    \item Passive sensing for non-cooperative UAVs
\end{itemize}
\\ \cline{1-6}

PRS & \begin{itemize}[leftmargin=*]
    \item High sensing accuracy potential
    \item Flexible resource configuration in time/frequency domains
\end{itemize}
&
High & Medium & High & \begin{itemize}[leftmargin=*]
    \item Ranging and Doppler estimation
    \item Periodic tracking
    \item Suitable when dedicated sensing resources can be scheduled
\end{itemize}
\\ \cline{1-6}

DMRS & \begin{itemize}[leftmargin=*]
    \item Medium-to-high sensing accuracy potential
    \item Strongly dependent on allocated bandwidth, pilot density, and temporal continuity
\end{itemize}
&
High & Low & Very high & \begin{itemize}[leftmargin=*]
    \item Continuous tracking during communication
    \item Applicable when the UAV is in connected mode
\end{itemize}
\\ \cline{1-6}

Data-assisted & \begin{itemize}[leftmargin=*]
    \item Highest sensing potential by exploiting both pilots and payload data
    \item Can approach the fundamental sensing performance bound under reliable data recovery
\end{itemize}
&
High & Low extra signaling overhead but high processing complexity & Medium & \begin{itemize}[leftmargin=*]
    \item High-precision tracking
    \item Requires an active communication link
\end{itemize}
\\

\bottomrule[1pt]
\end{tabular}

\end{threeparttable}
\end{table*}

\subsection{ISAC in Mixed Near- and Far-Field Channel}
In UAV surveillance, the wireless channel model depends strongly on the target trajectory and operating environment. UAVs monitored at long distances in open areas are often well approximated by the far-field model, whereas those flying through urban canyons or passing close to rooftop-mounted or large-aperture arrays may enter the near-field regime. Consequently, the sensing network should be designed to accommodate mixed near-field and far-field conditions over time and across different sensing BS.

For conventional far-field ISAC, extensive studies have investigated sensing performance metrics \cite{liu2021cramer, ren2023fundamental}, beamforming design \cite{zhao2022joint}, and the fundamental sensing-communication performance tradeoff \cite{chen2022generalized}. When the sensing target enters the near-field region, the inherent range-phase coupling in the spherical wavefront results in the performance metrics and sensing algorithms developed for the far-field regime are no longer effective. Consequently, a number of recent works have been devoted to near-field ISAC, proposing new performance metrics and sensing designs tailored to the near-field scenario \cite{wang2023near, zhao2024modeling, hua2025near}. Although a near-field model can in principle accommodate both near-field and far-field targets, the complex parameter dimensionality and limited gains on far-field links, together with the modeling errors and ISAC performance loss caused by mixed-field scatterers under mismatched models, highlight the need for efficient mixed-field ISAC frameworks that can flexibly capture both near-field and far-field components.

In \cite{xiao2024integrated}, the authors investigated transmit beamforming for mixed near-field and far-field ISAC in a bistatic system, where a multi-antenna BS serves multiple communication users while sensing a target that may reside in a different propagation region from the users. In the proposed framework, near-field and far-field channels are described by spherical-wave and plane-wave models, respectively, and a beamforming problem is formulated to maximize the sensing signal-to-clutter-plus-noise ratio (SCNR) under per-user SINR and total transmit-power constraints, which is tackled via semidefinite relaxation and an iterative optimization procedure. Numerical results show that conventional far-field-only beamforming suffers noticeable performance loss in mixed-field scenarios, whereas the mixed-field design achieves higher SCNR, especially with larger antenna arrays or relaxed communication SINR requirements and naturally reduces to pure near-field-only and far-field-only ISAC as special cases. The authors in \cite{wei2021channel} studied channel estimation for extremely large-scale massive MIMO under a more realistic hybrid-field propagation environment, where some scatters lie in the far field while others are in the near field. A hybrid-field channel model is introduced that simultaneously includes far-field and near-field path components, controlled by an adjustable parameter so that conventional far-field-only and near-field-only models appear as special cases. Building on this model, a hybrid-field orthogonal matching pursuit (OMP) algorithm is developed that separately estimates far-field and near-field components using different transform matrices and then combines them to recover the full channel, achieving lower normalized MSE than existing far-field and near-field OMP schemes for the same pilot overhead. The use of extremely large-scale arrays in mixed near- and far-field channels is examined in \cite{zhang2023mixed}, where a BS equipped with a very large linear array simultaneously serves one user in the near field and another in the far field. A closed-form expression for the interference power experienced by the near-field user due to the far-field beam is derived using Fresnel integrals and is subsequently employed to study how this interference scales with the number of antennas, user angles, and the near-field user's distance. The analysis shows that strong interference can occur when the array size and near-field distance are relatively small. Building on this interference characterization, an explicit rate-loss expression for the near-field user is obtained and validated through simulations, demonstrating that substantial rate degradation may arise in mixed-field scenarios and highlighting the need for more advanced digital beamforming strategies to better suppress inter-user interference.

Xiao \textit{et al.} studied transmit beamforming for integrated sensing and communication in a mixed near- and far-field scenario with multiple targets and multiple users \cite{xiao2025mixed}. A dual-functional base station (DFBS) equipped with separate transmit and receive arrays simultaneously serves near- and far-field communication users and detects near- and far-field targets. Two beamforming formulations are considered: a sensing-centric design that maximizes the minimum SCNR across all targets subject to user SINR and power constraints, and a communication-centric design that maximizes the minimum user SINR while guaranteeing SCNR requirements for all targets under the same power budget. Both problems are tackled via semidefinite relaxation combined with a Dinkelbach-based successive convex approximation procedure. Numerical results reveal a clear tradeoff between sensing and communication performance in mixed-field environments. The results also show how antenna size and transmit power impact this tradeoff, and demonstrate that mixed-field beamforming based on mixed field channel models outperforms far-field-only benchmarks and can approach near-field performance with notably lower computational overhead.

\section{MIMO OFDM-Enabled UAV Surveillance Techniques}
\label{sec3}
\begin{figure*}[hbt]
\centering
\includegraphics[scale=0.6]{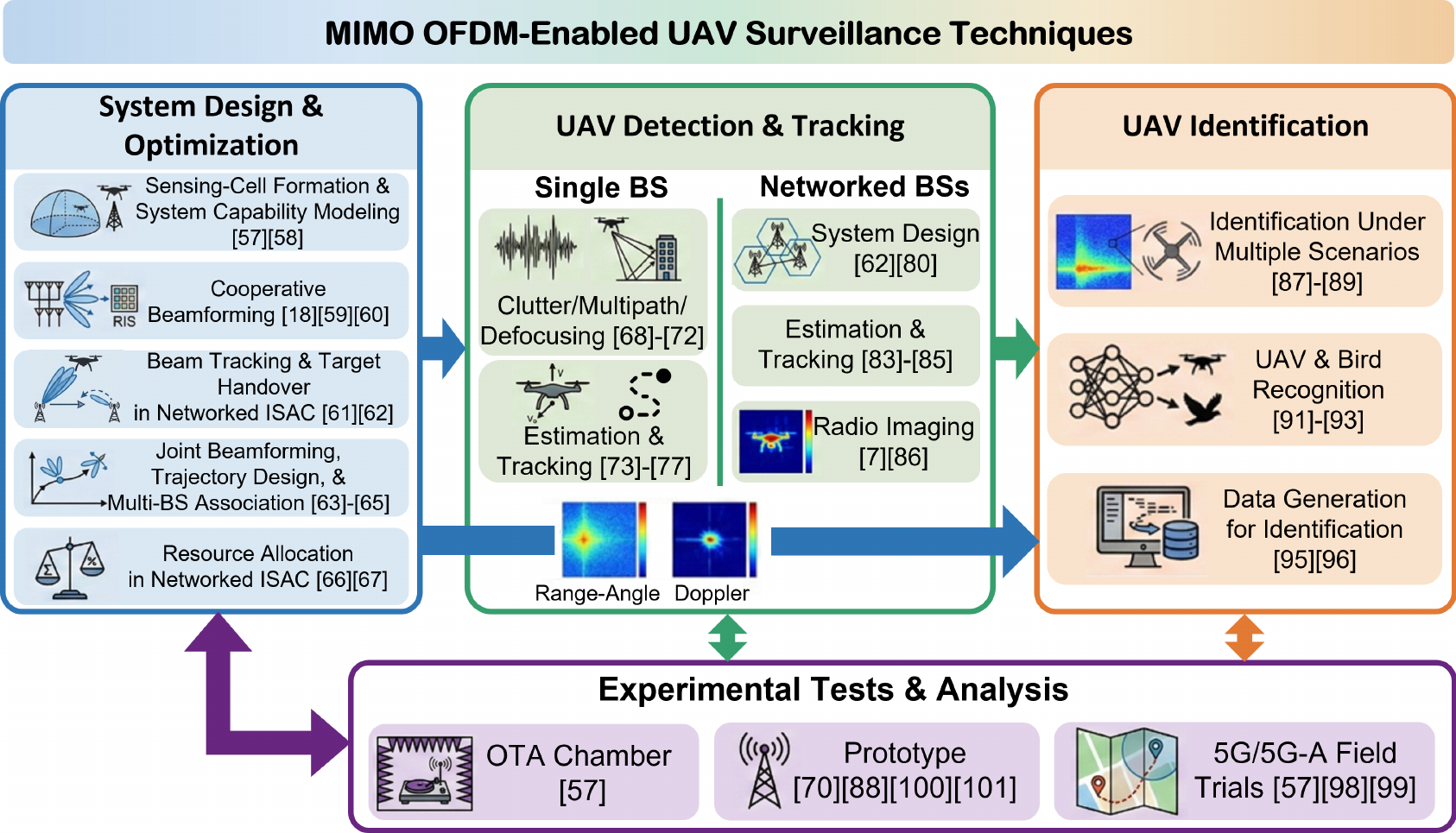}
\caption{MIMO OFDM-enabled UAV surveillance techniques.}
\label{fig:5}
\end{figure*}

In this section, we focus on how MIMO OFDM-enabled ISAC is applied to low-altitude non-cooperative UAV surveillance. First, we review the system design and optimization of ISAC networks. We then survey various algorithms developed for UAV detection and tracking. Next, we discuss UAV identification techniques. Finally, we summarize recent experimental tests and field trials that demonstrate the feasibility of UAV surveillance using practical MIMO OFDM ISAC platforms. Related works on MIMO OFDM-enabled UAV surveillance techniques are shown in Fig. \ref{fig:5}.

\begin{figure*}[hbt]
\centering
\includegraphics[scale=0.55]{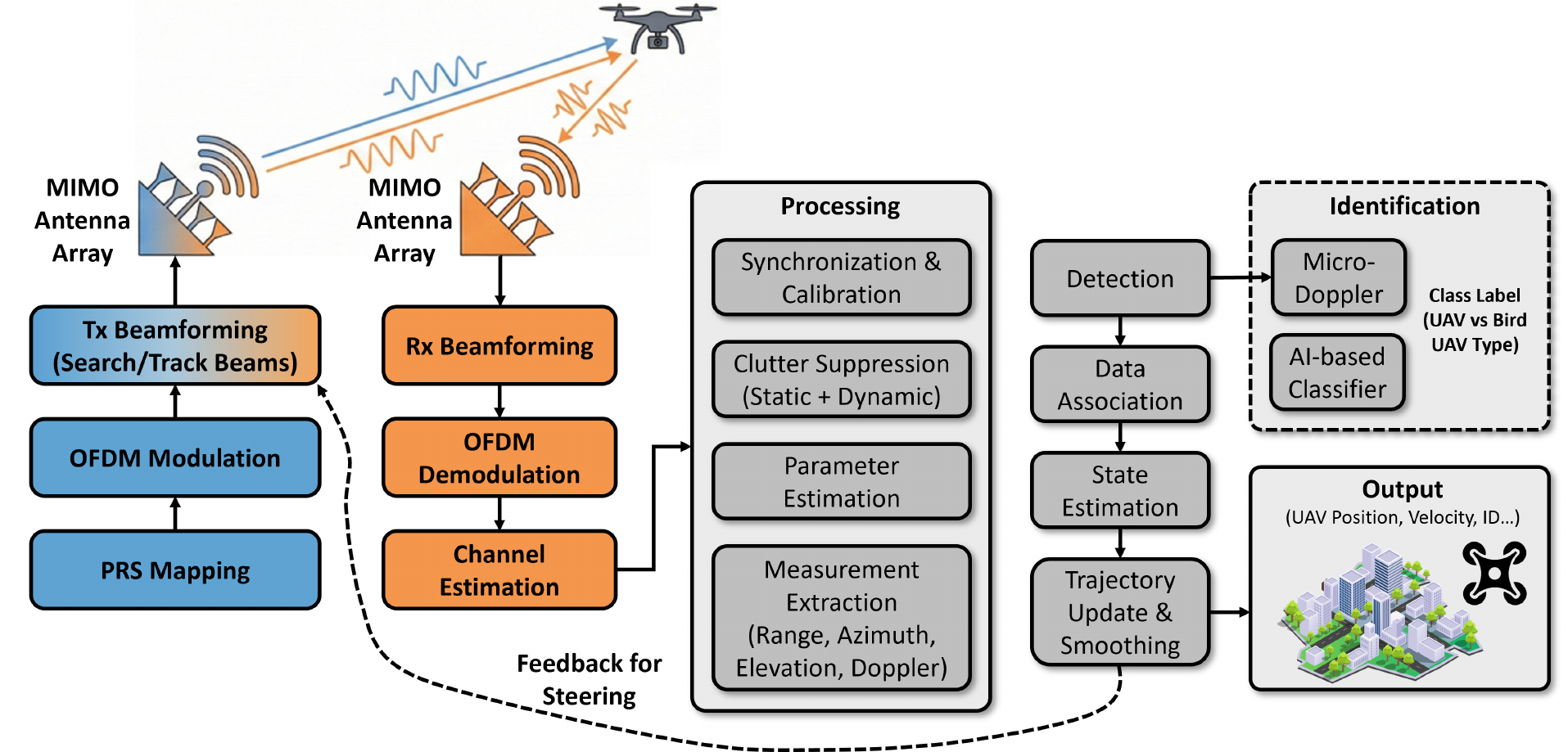}
\caption{A general MIMO OFDM-based ISAC framework for low-altitude UAV surveillance.}
\label{fig:6}
\end{figure*}

To provide a unified perspective, Fig. \ref{fig:6} abstracts a general MIMO OFDM-based ISAC framework for low-altitude UAV surveillance. Rather than representing a specific system implementation, this figure offers a high-level view of the common signal processing and inference pipeline.

In this framework, OFDM signaling with sensing-oriented reference signals (e.g., PRS) is used as a representative example. At the transmitter, sensing-enabled OFDM waveforms are generated and steered via MIMO beamforming for target search and tracking. At the receiver, beamforming, OFDM demodulation, and channel estimation are performed to extract sensing-related measurements, which are further processed through synchronization, clutter suppression, and parameter estimation to obtain range, angle, and Doppler information. Based on these intermediate measurements, detection, data association, state estimation, and trajectory smoothing are carried out, followed by micro-Doppler analysis and AI-based classification for UAV identification. In addition, feedback from tracking results can be exploited to adaptively refine beamforming strategies. Overall, the framework offers a unified abstraction that bridges algorithm-level developments and practical ISAC system implementations.

The following subsections review representative research works corresponding to each component of this framework.

\subsection{System Design and Optimization}
This subsection reviews key system design and optimization approaches for low-altitude ISAC from a network-level perspective. In contrast to conventional link-level optimization, UAV surveillance requires coordinated multi-BS sensing, efficient communication-sensing resource sharing, and scalable system design to support wide-area and persistent monitoring. A summary of the related works in this subsection is shown in Table \ref{tab:3}.

\begin{table*}[!t]
\renewcommand{\arraystretch}{1.3}
\centering
\caption{Summary of ISAC System Modeling and Optimization for UAV Surveillance.}
\label{tab:3}
\begin{threeparttable}

% 关键：m{} 列格式让内容垂直居中
\begin{tabular}{
    >{\centering\arraybackslash}m{4cm}|
    >{\centering\arraybackslash}m{2.8cm}|
    >{\centering\arraybackslash}m{9.2cm}
}
\toprule[1pt]
\textbf{Category} & \textbf{Reference} & \textbf{Key Contributions} \\
\midrule

% =============================
% System Capability Modeling
% =============================
\multirow{2}{4cm}{%
    \centering
    \textbf{\makecell{Sensing-Cell Formation \\ \& System Capability Modeling}}%
    \vspace{-0.8\baselineskip}%
}
& Li \textit{et al.} \cite{li2025low} &
\begin{itemize}[leftmargin=*]
  \item Develops a 3D single-BS sensing model that clarifies sensing boundaries and key design parameters.
  \item Proposes a multi-BS complementary coverage scheme and 3D cellular-like topology for seamless low-altitude sensing.
\end{itemize}
\\ \cline{2-3}
& Wan \textit{et al.} \cite{wan2025sensing} &
\begin{itemize}[leftmargin=*]
  \item Defines sensing capacity as a new ISAC performance metric.
  \item Establishes closed-form relationships between sensing capacity, SNR, and PD.
\end{itemize}
\\
\midrule

% =============================
% Beamforming Design
% =============================
\multirow{3}{4cm}{\centering \textbf{Cooperative Beamforming \\ in Networked ISAC}\vspace{-8\baselineskip}}
& Zhang \textit{et al.} \cite{zhang2024cooperative} &
\begin{itemize}[leftmargin=*]
  \item Jointly optimizes multi-cell ISAC transceiver beamforming to maximize sensing SCNR under downlink SINR and power constraints.
  \item Presents a centralized AO/successive convex approximation (SCA)/Dinkelbach beamforming method that boosts sensing SCNR while meeting SINR and power limits.
  \item Proposes a distributed beamforming algorithm via primal decomposition, achieving low-overhead multicell coordination.
\end{itemize}
\\ \cline{2-3}

& Wang \textit{et al.} \cite{wang2025coordinated} &
\begin{itemize}[leftmargin=*]
  \item Proposes a coordinated DFBS-RIS beamforming framework for ISAC that maximizes communication sum-rate while satisfying sensing SNR constraints.
  \item Develops a FP- and AO-based algorithm that decomposes the non-convex joint beamforming problem.
\end{itemize}
\\ \cline{2-3}

& Zhou \textit{et al.} \cite{zhou2025full} &
\begin{itemize}[leftmargin=*]
  \item Proposes a zero-overhead OFDM-ISAC framework that removes dedicated sensing time-frequency resources to enhance both communication and sensing.
  \item Proposes a power-efficient joint communication-sensing beamforming design for the searching stage.
  \item Develops a low-complexity two-stage super-resolution sensing method with spatial denoising to enhance sensing accuracy under low SNR.
\end{itemize}
\\
\midrule

\multirow{2}{4cm}{\centering \textbf{Beam Tracking \& Target Handover \\ in Networked ISAC}\vspace{-2.0\baselineskip}}
& Yang \textit{et al.} \cite{yang2025cooperative1} &
\begin{itemize}[leftmargin=*]
  \item Propose a multi-BS cooperative sensing framework with EKF-based state prediction.
  \item Develop predictive beamforming for motion-aware beam alignment.
\end{itemize}
\\ \cline{2-3}

& Feng \textit{et al.} \cite{feng2025networked} &
\begin{itemize}[leftmargin=*]
  \item Propose a networked ISAC tracking framework with virtual sensing cells.
  \item Develop seamless handover via dynamic BS selection and sensing cell transition.
\end{itemize}
\\
\midrule

% =============================
% Beamforming & Trajectory
% =============================
\multirow{2}{4cm}{%
    \centering
    \textbf{\makecell{Joint Beamforming, \\ Trajectory Design, \\ \& Multi-BS Association}}%
    \vspace{-3\baselineskip}%
}
& Wang \textit{et al.} \cite{wang2024isac} &
\begin{itemize}[leftmargin=*]
  \item Proposes a cooperative ISAC framework where EKF-based sensing fusion between the BS and UAV enhances detection accuracy and enables beyond line of sight (LoS) sensing.
  \item Develops a joint beamforming and UAV trajectory optimization algorithm to maximize communication rate under sensing constraints.
\end{itemize}
\\ \cline{2-3}

& Cheng \textit{et al.} \cite{cheng2024networked,cheng2025networked} &
\begin{itemize}[leftmargin=*]
  \item Jointly optimizes coordinated GBS beamforming, UAV trajectory, and GBS-UAV association under power, flight, and 3D sensing illumination constraints.
  \item Develops an AO/semi-definite relaxation (SDR)/SCA algorithm to solve the mixed-integer coupled problem, achieving efficient sensing-communication tradeoff improvements.
\end{itemize}
\\
\midrule

% =============================
% Resource Allocation
% =============================
\multirow{2}{4cm}{\centering \textbf{Resource Allocation \\ in Networked ISAC}\vspace{-5\baselineskip}}
& Li \textit{et al.} \cite{li2025joint} &
\begin{itemize}[leftmargin=*]
  \item Formulates a unified resource allocation problem that jointly optimizes UAV assignment and transceiver beamforming to maximize the minimum sensing SINR.
  \item Develops an efficient iterative solution combining closed-form receive beamforming and fractional transmit beamforming optimization to solve the mixed-integer problem.
\end{itemize}
\\ \cline{2-3}

& Gao \textit{et al.} \cite{gao2024collaborative} &
\begin{itemize}[leftmargin=*]
  \item Proposes a multi-beam power allocation scheme for multi-BS collaborative ISAC.
  \item Formulates a utility-based optimization using SINR and CRLB for joint communication-sensing performance.
  \item Develops the MCBPA algorithm for approximate optimal beam power allocation.
\end{itemize}
\\

\bottomrule[1pt]
\end{tabular}

\end{threeparttable}
\end{table*}

\subsubsection{Sensing-Cell Formation and System Capability Modeling}
The characterization of sensing coverage and system capability forms the foundation of network-level ISAC design, determining how sensing cells are formed and coordinated across multiple BSs.

Li \textit{et al.} presented a comprehensive system-level study on low-altitude ISAC network design, focusing on the sensing capability modeling of single BS and the construction of a 3D cellular-like multi-BS topology \cite{li2025low}. The article analyzes the influence of the antenna's vertical field of view (FOV) and highlights the differences in cell radius and beam characteristics between communication and sensing modes. Compared with prior work, the proposed method provides a more precise characterization of the sensing boundaries and effective coverage range of a single BS. In addition, the authors introduce a multi-BS 3D network topology that extends conventional 2D cellular structures into the vertical domain, enabling seamless coverage across different altitude layers. The framework is validated through both far-field over-the-air (OTA) tests in a controlled mmWave anechoic chamber and real-world field trials on a commercial mmWave ISAC network, demonstrating stable sensing performance and confirming the feasibility of the proposed deployment strategy. This work provides practical guidance for designing low-altitude ISAC BSs and network architectures.

Wan \textit{et al.} analyzed the sensing performance limits of a single-BS ISAC system for low-altitude UAV surveillance \cite{wan2025sensing}. To address the challenges of detecting unauthorized UAVs, they introduced a novel metric termed sensing capacity, defined as the maximum number of UAVs that can be simultaneously detected. Using SNR and probability of detection (PD) as intermediate variables, this paper derives a closed-form expression for the maximum detectable UAV count under an SNR constraint, and further proposes an approximate solution for determining this limit under PD constraints. The findings offer a useful system-level guideline for ISAC BS configuration and deployment.

\subsubsection{Cooperative Beamforming in Networked ISAC}
To improve sensing performance in networked ISAC, multiple BSs need to cooperatively illuminate targets, which introduces the need for coordinated beam control across the network.

Zhang \textit{et al.} studied cooperative transceiver beamforming for multi-cell anti-UAV ISAC systems \cite{zhang2024cooperative}, where multiple BSs collaboratively conduct UAV sensing. The authors jointly optimize the ISAC transmit and receive beamformers at BSs and downlink users to maximize the SCNR, while satisfying communication requirements and power constraints. A centralized algorithm is developed using alternating optimization, successive convex approximation, and Dinkelbach methods, and a distributed algorithm is further proposed via primal decomposition to decouple inter-cell interference and reduce backhaul overhead. Results show significant sensing performance improvements over single BS sensing and demonstrate effective suppression of interference and clutter in multi-cell ISAC settings.

Wang \textit{et al.} proposed a coordinated beamforming framework for RIS-empowered ISAC systems over secure low-altitude networks \cite{wang2025coordinated}. The dual-functional base station (DFBS) simultaneously provides communication services for legitimate UAVs and detects unauthorized UAVs. By jointly optimizing the active precoding at the DFBS and the passive beamforming at the RIS via fractional programming (FP) and alternating optimization, the method maximizes the communication sum-rate under sensing SNR and power constraints. Results show that the joint RIS-aided design delivers significantly higher data rates than disjoint schemes, highlighting the essential role of RIS in balancing communication performance and sensing capability.

Zhou \textit{et al.} proposed a MIMO OFDM ISAC framework tailored for low-altitude UAV applications, eliminating the need for dedicated sensing time-frequency resources and achieving zero sensing overhead through a jointly designed transmit beamforming scheme that simultaneously satisfies communication and sensing requirements \cite{zhou2025full}. The proposed approach improves the communication sum rate while enhancing sensing resolution, unambiguous range, and estimation accuracy. To facilitate practical deployment, a low-complexity search beamforming strategy and a two-stage super-resolution sensing algorithm are further developed. Simulation results confirm that the proposed design consistently outperforms traditional ISAC schemes in both communication sum rate and sensing quality, highlighting its strong potential for future ISAC networks serving low-altitude UAVs.

\subsubsection{Beam Tracking and Target Handover in Networked ISAC}
In practical networked ISAC systems, UAV mobility introduces strong dynamics, requiring continuous beam alignment and seamless sensing coverage across multiple BSs. This motivates the integration of beam tracking and target handover mechanisms to maintain sensing continuity in dynamic environments.

To address beam misalignment caused by UAV mobility, recent works have investigated predictive beam tracking schemes that exploit sensing signals to estimate and predict target states for motion-aware beamforming design. For example, Yang \textit{et al.} proposed a cooperative sensing-assisted predictive beam tracking framework for MIMO-OFDM networked ISAC systems \cite{yang2025cooperative1}. In this work, multiple BSs perform cooperative sensing by first conducting local target estimation based on echo signals and then fusing multi-BS measurements via an EKF to predict the target state, including position and velocity. Based on the predicted target parameters, a predictive beamforming design is formulated to maximize communication rate subject to sensing accuracy constraints in the next time slot. The results demonstrate that multi-BS cooperative sensing significantly improves tracking accuracy, which in turn facilitates more reliable beam alignment compared with conventional single-BS or non-predictive schemes. Notably, the beam tracking process is performed over multiple tracking time slots, allowing continuous adaptation of beam directions to dynamic UAV motion.

In parallel, target handover has been investigated to maintain sensing continuity when UAVs move across different BS coverage regions. Feng \textit{et al.} proposed a networked ISAC-based UAV tracking and handover framework built upon the concept of virtual sensing cells (VSC) \cite{feng2025networked}. Within each VSC, a primary BS is dynamically selected for sensing transmission, while multiple BSs cooperatively receive echoes, enabling flexible sensing coverage adaptation. To address UAV mobility across cells, both primary BS handover and VSC handover strategies are developed. The primary BS handover dynamically switches the transmitting BS based on signal quality and blockage conditions, while the VSC handover enables seamless transition of sensing responsibility when UAVs move between neighboring sensing cells. This design ensures continuous tracking without interruption and enhances robustness against blockage and inter-cell interference in dynamic low-altitude environments.

\subsubsection{Joint Beamforming, Trajectory Design, and Multi-BS Association}
In dynamic UAV scenarios, system performance is further influenced by UAV mobility and the need for adaptive association across multiple BSs, which introduces coupling between beamforming, trajectory design, and network-level coordination.

Wang \textit{et al.} studied a cellular-connected ISAC UAV system that performs cooperative detection via BS-UAV sensing data fusion \cite{wang2024isac}. An extended Kalman filter (EKF)-based fusion algorithm is first developed to combine BS and UAV measurements. Based on the fused results, a joint beamforming and trajectory design problem is formulated to maximize the communication rate under power and sensing constraints, addressing UAV mobility and energy limitations. The non-convex problem is efficiently solved via successive convex approximation and semidefinite relaxation, achieving over 31\% improvement in data rate.

Cheng \textit{et al.} proposed a networked ISAC framework for low-altitude UAV operations \cite{cheng2024networked, cheng2025networked}. Multiple ground base stations (GBSs) jointly optimize coordinated transmit beamforming, the authorized UAVs' trajectories, and their GBS associations to support communication while simultaneously sensing non-cooperative targets. The resulting non-convex optimization problem is addressed using alternating optimization, successive convex approximation, and semidefinite relaxation. Simulation results demonstrate clear improvements in the communication-sensing tradeoff, with horizontally deployed antenna arrays and interference-cancellation-capable UAV receivers offering additional performance gains.

\subsubsection{Resource Allocation in Networked ISAC}

Efficient allocation of sensing and communication resources is essential for scalable networked ISAC systems, especially when multiple BSs and multiple targets coexist.

Li \textit{et al.} proposed a fairness-oriented resource allocation framework for networked ISAC, tailored to fast-moving UAV detection under strong and rapidly varying interference \cite{li2025joint}. The study jointly optimizes target allocation and ISAC beamforming across multiple BSs to maximize the minimum radar sensing SINR, while ensuring communication quality of service (QoS) and BS power constraints. An AO-based method is developed, where receive beamforming is derived via the generalized Rayleigh quotient and transmit beamforming is solved using Dinkelbach's method. Simulation results demonstrate that the proposed design provides robust and equitable sensing performance.

Gao \textit{et al.} proposed a multi-BS collaborative ISAC framework for non-cooperative UAV localization, where multiple BSs jointly allocate multi-beam transmit power to meet differentiated sensing QoS requirements \cite{gao2024collaborative}. They introduce a multi-BS collaborative beam power allocation (MCBPA) algorithm that optimizes the CRLB of UAV location estimation via an exponential utility formulation. Simulation results show that MCBPA reduces sensing error for targets by 51\% and saves about 20\% transmit power compared with single-BS average beam power allocation.

\subsection{UAV Detection and Tracking}
The detection and tracking of UAVs using ISAC represents a rapidly growing research domain. This paper will provide a comprehensive review of existing studies from two viewpoints: single BS and networked BSs.
\subsubsection{Single BS}
Detecting non-cooperative UAVs under complex conditions is highly challenging. To tackle this problem, researchers have carried out various related investigations. Studies on UAV detection and tracking with a single BS are summarized in Table \ref{tab:4}.

\begin{table*}[!t]
\renewcommand{\arraystretch}{1.3}
\centering
\caption{Summary of UAV Detection and Tracking Using Single-BS ISAC.}
\label{tab:4}
\begin{threeparttable}

% 关键：m{} 列格式让内容垂直居中
\begin{tabular}{
    >{\centering\arraybackslash}m{3cm}|
    >{\centering\arraybackslash}m{2.5cm}|
    >{\centering\arraybackslash}m{5.5cm}|
    >{\centering\arraybackslash}m{5.0cm}
}
\toprule[1pt]
\textbf{Problems} & \textbf{References} & \textbf{Algorithms} & \textbf{Key Results} \\
\midrule

% =============================
% System Capability Modeling
% =============================
\multirow{2}{3cm}{%
    \centering
    \textbf{\makecell{Detection in \\ Clutter Environments}}%
    \vspace{-5\baselineskip}%
}
& Luo \textit{et al.} \cite{luo2024integrated} &
\begin{itemize}[leftmargin=*]
  \item Mixed clutter-target channel model.
  \item Static clutter filtering.
  \item ADSE and MSJD for detection and angle estimation.
  \item Extended subspace method for range and velocity estimation.
\end{itemize} &
\begin{itemize}[leftmargin=*]
  \item Robust detection under clutter.
  \item Outperforms methods ignoring clutter.
\end{itemize}
\\ \cline{2-4}
& Blandino  \textit{et al.} \cite{blandino2025detecting} &
\begin{itemize}[leftmargin=*]
  \item PRS-based monostatic sensing pipeline.
  \item Temporal averaging for static clutter suppression.
  \item Angle and range extraction with peak-to-average ratio (PAR)-based detection.
\end{itemize}&
\begin{itemize}[leftmargin=*]
  \item In UMi, miss detection reaches 16\% at 25 m and drops to about 3\% at 50 m, becoming negligible above 100 m.
  \item In UMa, miss detection remains near zero below 100 m and rises slightly to about 1\% at 200 m due to reduced SNR at longer BS-ST distances.
\end{itemize}

\\
\midrule

% =============================
% Beamforming Design
% =============================
\multirow{2}{3cm}{%
    \centering
    \textbf{\makecell{Detection in \\ Multipath Environments}}%
    \vspace{-5\baselineskip}%
}
& Su \textit{et al.} \cite{su2024integrated} &
\begin{itemize}[leftmargin=*]
  \item UAV multipath reflection wave model.
  \item Reinforcement learning-based adaptive M/N detector.
  \item Minimum SNR reward mechanism with iterative policy update.
\end{itemize}&
\begin{itemize}[leftmargin=*]
  \item Target detection success rate of 91.2\%.
\end{itemize}
\\ \cline{2-4}

& Yin \textit{et al.} \cite{yin2025deep} &
\begin{itemize}[leftmargin=*]
  \item Micro-Doppler spectrogram extraction via STFT for each path.
  \item ResNet-based feature extraction of individual UAV micro-Doppler signatures.
  \item Signature clustering (k-means + hierarchical) to associate multipaths from the same UAV.
\end{itemize}&
\begin{itemize}[leftmargin=*]
  \item False alarm rate reduced by 86\% compared to baseline.
  \item Missed detection rate increases slightly to about 6\%, but remains low.
  \item Localization accuracy maintained.

\end{itemize}
\\
\midrule

% =============================
% Beamforming & Trajectory
% =============================

\textbf{\makecell{Detection under \\ Motion-induced \\ Defocusing}}
& Gui \textit{et al.} \cite{gui2025mmw} &
\begin{itemize}[leftmargin=*]
  \item Far-field wavenumber-domain ISAR imaging.
  \item Minimum entropy-based motion parameter formulation.
  \item PSO-based optimization to estimate target velocity.
\end{itemize}&
\begin{itemize}[leftmargin=*]
  \item Effectively mitigates defocusing caused by unknown target motion.
  \item Produces well-focused multi-frame ISAR images for vehicles and UAVs.
\end{itemize}
\\

\midrule

% =============================
% Resource Allocation
% =============================
\multirow{2}{3cm}{\centering \textbf{Parameter Estimation}\vspace{-5\baselineskip}}
& Lin \textit{et al.} \cite{lin2025isac} &
\begin{itemize}[leftmargin=*]
  \item Near-field ISAC signal model and MCRB-based analysis of mobility-induced model misspecification.
  \item Two-stage localization: tensor-decomposition initialization and quasi-Newton refinement for joint position/velocity estimation.
\end{itemize}&
\begin{itemize}[leftmargin=*]
  \item Enables better near-field high-mobility UAV parameter estimation and localization.
\end{itemize}
\\ \cline{2-4}

& Luo \textit{et al.} \cite{luo20256d} &
\begin{itemize}[leftmargin=*]
  \item ESPRIT-based estimation of horizontal angle, pitch angle, distance, and per-antenna virtual velocity extraction.
  \item Plane fitting to jointly recover radial velocity, horizontal and pitch angular velocities.
\end{itemize}&
\begin{itemize}[leftmargin=*]
  \item Achieves full 6D motion parameter estimation using a single massive-MIMO BS.
\end{itemize}
\\

\midrule

% =============================
% Resource Allocation
% =============================
\multirow{2}{3cm}{\centering \textbf{Tracking}\vspace{-8\baselineskip}}
& Yan \textit{et al.} \cite{yan2025uav, yan2025uav1} &
\begin{itemize}[leftmargin=*]
  \item Beam scanning and clutter suppression for real-time target detection.
  \item PRDFT-based motion parameter estimation.
  \item WGVDNN trajectory association.
  \item IMMUKF-based trajectory prediction.
\end{itemize}&
\begin{itemize}[leftmargin=*]
  \item Supports real-time discovery of new targets as well as continuous tracking of previously detected ones.
\end{itemize}
\\ \cline{2-4}

& Cui \textit{et al.} \cite{cui2025research} &
\begin{itemize}[leftmargin=*]
  \item Spatial rotation-based parameter estimation.
  \item	Cubic interpolation for short-term occlusion. UKF prediction (constant velocity) for long-term occlusion.
\end{itemize}&
\begin{itemize}[leftmargin=*]
  \item Avg. position error: 0.68-0.82 m (no occlusion), 1.05-2.18 m (short), 1.85-3.79 m (long).
  \item Error convergence occurs within 1 s for short occlusion and within 5 s for long occlusion.
  \item Velocity error converges to less than 0.2 m/s after the target becomes visible again.
\end{itemize}
\\

\bottomrule[1pt]
\end{tabular}

\end{threeparttable}
\end{table*}

Luo \textit{et al.} explored an ISAC framework for sensing dynamic targets in cluttered environments while maintaining reliable user communications \cite{luo2024integrated}. The BS simultaneously maintains communications via multiple fixed communication beams while employing a rotating sensing beam to achieve full spatial coverage. To mitigate mutual interference, the service area is divided into sensing-beam-for-sensing (S4S) and communication-beam-for-sensing sectors (C4S), with dedicated beamforming and power allocation strategies for each. A mixed sensing channel model incorporating both static environmental clutter and dynamic target echoes is built, enabling clutter suppression by exploiting Doppler distinctions between static and moving objects. After clutter removal, angle-Doppler spectrum estimation (ADSE) and joint detection over multiple subcarriers (MSJD) are applied for dynamic target detection and angle estimation, while distance and velocity are obtained separately using an extended subspace algorithm. Simulation results demonstrate that the proposed method can enhance detection robustness and sensing accuracy under cluttered conditions. Blandino \textit{et al.} investigated the feasibility of employing standardized 5G NR PRS for monostatic radar sensing of UAVs in urban environments \cite{blandino2025detecting}. A complete 5G NR-based radar processing chain is provided, encompassing clutter suppression, angle and range estimation, and 3D position reconstruction. Simulation results show that in the Urban Micro (UMi) scenario, dense urban clutter causes significant performance degradation, with miss detection rates reaching 16\% at 25 m and decreasing to about 3\% at 50 m, becoming negligible above 100 m. In contrast, the Urban Macro (UMa) scenario exhibits almost no miss detection below 100 m due to the elevated BS placement and reduced clutter, with only a slight increase (around 1\%) at 200 m caused by the lower SNR at longer distances. These results indicate that monostatic 5G NR radar-based detection is feasible but highly sensitive to clutter conditions.

Su \textit{et al.} developed a reinforcement learning-based UAV detection method \cite{su2024integrated}. This paper first builds a reflection wave model for UAVs in multipath environments and then introduces an adaptive M/N detector. The detector formulates a reinforcement learning model using the minimum SNR as the reward and iteratively updates the parameter selection policy without relying on human experience. Experimental results show that the proposed method achieves a target detection success rate of 91.2\%. Yin \textit{et al.} proposed a deep learning-assisted ISAC framework for UAV detection in multipath environments \cite{yin2025deep}. In urban canyons, the UAV produces both direct and indirect reflections, which are often misinterpreted as multiple targets, resulting in high false alarm rates. The proposed method employs ResNet \cite{he2016deep} to extract micro-Doppler signatures from each path. Then, a clustering algorithm associates paths belonging to the same UAV and estimates the total number of UAVs. This approach effectively distinguishes multipaths of different UAVs, thereby reducing false alarms and enhancing overall detection accuracy. Gui \textit{et al.} investigated a mmWave ISAC imaging framework for non-cooperative moving targets, using an inverse synthetic aperture radar (ISAR) architecture integrated with 5G OFDM communication signals \cite{gui2025mmw}. To address the challenges posed by unknown target motion and outdoor sensing conditions, a far-field wavenumber-domain imaging algorithm using minimum image entropy is introduced to achieve robust multi-frame imaging. The method uses particle swarm optimization (PSO) to estimate target velocity and automatically compensate motion-induced defocusing. Experiments demonstrate that the proposed 5G mmWave ISAC system can reconstruct well-focused ISAR images for both ground vehicles and airborne UAVs.

For localization and tracking, Lin \textit{et al.} analyzed near-field ISAC localization for high-mobility UAVs, where spherical wave propagation introduces strong coupling among range, angle, and Doppler parameters \cite{lin2025isac}. This paper quantifies the performance degradation in target localization resulting from mobility-induced model mismatches through the misspecified Cram\'er-Rao bound (MCRB) and derives theoretical limits for achievable accuracy. To address these coupling effects, a two-stage localization algorithm is proposed: tensor decomposition is first employed for coarse estimation, followed by quasi-Newton refinement to jointly recover range, angle, and velocity parameters. Simulation results demonstrate that the proposed approach significantly improves localization and velocity estimation accuracy in near-field high-mobility scenarios, achieving performance close to the theoretical bounds. Luo \textit{et al.} investigated a monostatic MIMO OFDM ISAC framework capable of jointly estimating 6D motion parameters of dynamic targets \cite{luo20256d}. This paper first uses estimating signal parameters via rotational variation techniques (ESPRIT) \cite{roy2002esprit} to obtain the target's horizontal angle, pitch angle, distance, and the virtual velocities across the antenna array. The virtual velocities are then processed using a plane-fitting method to extract the remaining motion components, including radial and angular velocities. The results demonstrate that a single massive-MIMO ISAC BS can successfully recover the full set of 6D motion parameters.

Yan \textit{et al.} proposed an ISAC framework for UAV trajectory monitoring \cite{yan2025uav, yan2025uav1}. The system integrates beam scanning, clutter suppression, and a phase-rotated discrete Fourier transform (PRDFT) algorithm to estimate the motion parameters of dynamic targets. A wave gate and velocity difference nearest neighbor (WGVDNN) association method, together with an interactive multiple-model unscented Kalman filter (IMMUKF), enables accurate trajectory prediction and continuous target monitoring. Simulation results demonstrate that the proposed approach supports real-time discovery of new targets as well as continuous tracking of previously detected ones. Cui \textit{et al.} proposed a UAV tracking method with 6G ISAC \cite{cui2025research}. The system employs a dual-functional transceiver that simultaneously transmits communication and sensing beams via hybrid beamforming and receives sensing echoes using all-digital beamforming. To address LoS interruptions, the authors introduce an unscented Kalman filter (UKF) based tracking scheme that incorporates cubic interpolation and a constant velocity occlusion model for motion prediction. Simulation results show that: (1) without occlusion, the average X/Y/Z position errors are 0.82 m, 0.79 m, and 0.68 m; (2) under short-term occlusion, the errors increase to 1.25 m, 2.18 m, and 1.05 m, with a 1 s convergence time after LoS recovery; (3) under long-term occlusion, the errors rise to 2.87 m, 3.79 m, and 1.85 m, requiring 5 s to reconverge; and (4) the velocity error rapidly decreases to within 0.2 m/s once observations resume.

\subsubsection{Networked BSs}
To overcome the limitations of single-BS detection and tracking, researchers leverage networked BSs to enable UAV sensing. Studies on UAV detection and tracking with networked BSs are summarized in Table \ref{tab:5}.

Chen \textit{et al.} investigated an Internet of Things (IoT)-enabled multi-BS ISAC framework that transforms cellular BSs into cooperative, edge-intelligent sensing nodes for real-time UAV surveillance \cite{chen2025iot}. The system adopts a four-layer architecture, terminal, edge, IoT platform, and cloud, supporting real-time exchange of echoes and low-level features and enabling spatial and temporal coordination via adaptive beam registration and cross-correlation timing. A hierarchical processing pipeline integrates coarse data-level processing with edge-assisted refinements, achieving low latency and scalable multi-point sensing. Simulation and preliminary hardware results show high detection reliability (85-90\%), sub-frame latency (15-25 ms), and 10-20\% potential energy-efficiency gains, illustrating its feasibility for smart-city airspace safety and cooperative UAV monitoring. Feng \textit{et al.} explored a networked ISAC architecture for UAV tracking and handover in the low-altitude economy \cite{feng2025networked}. Within each VSC, the BSs mitigate static clutter \cite{shrader1970mti} and extract local angle and range information using the multiple signal classification (MUSIC)-based sensing module, providing structured low-level measurements for networked perception. These measurements are then aggregated by a centralized EKF fusion engine, which maintains consistent and reliable multi-UAV tracking. To ensure persistent sensing coverage, the system incorporates both primary BS and VSC handover mechanisms, enabling seamless transition of sensing responsibilities as UAVs move across cells. Simulation results confirm the effectiveness of this architecture for cooperative UAV monitoring in urban low-altitude economy (LAE) environments.

Regarding parameter estimation, Li \textit{et al.} presented an early communication-based passive sensing framework that jointly estimates target position and velocity using OFDM signals \cite{li2019multi}. The proposed two-stage algorithm first formulates a sparsity-constrained nonconvex optimization problem for joint delay-Doppler estimation and demodulation error recovery, which is solved using conjugate gradient descent. Then, target positions and velocities are determined by solving a set of nonlinear equations and using a neural network. Although not a fully ISAC system, this work demonstrates the feasibility of reusing communication signals for sensing, serving as a precursor to later cooperative ISAC localization studies. Lu \textit{et al.} proposed a multi-BS cooperative ISAC method for UAV localization and velocity estimation \cite{lu2024integrated}. To overcome the low sensing accuracy of single BS, a MUSIC-based symbol-level fusion framework is developed, consisting of a single-BS preprocessing stage that superposes multiple spectral functions to enhance parameter accuracy, followed by a lattice-point searching step that determines the fusion result by minimizing the error relative to the preprocessing reference. Zhang \textit{et al.} proposed a cooperative ISAC framework compatible with 5G NR, where information-bearing OFDM communication signals are reused for sensing \cite{zhang2024target}. A two-stage scheme is investigated: in the first stage, delay-Doppler features are extracted through a 2D fast FFT to obtain bistatic ranges; in the second stage, target locations are estimated using a matching-based algorithm that eliminates ill-conditioned measurements, thereby achieving high-accuracy localization. Tang \textit{et al.} introduced a two-stage cooperative ISAC scheme for low-altitude UAV sensing \cite{tang2025cooperative}. In the first stage, each BS performs monostatic parameter estimation using a tensor decomposition model, where a spatial-smoothing tensor decomposition exploiting the Vandermonde structure and a reduced-dimensional angle of arrival (AoA) estimator based on the generalized Rayleigh quotient (GRQ) are applied to obtain the UAVs' angles, ranges, and Doppler parameters. In the second stage, multi-BS estimates are fused via a minimum-spanning-tree (MST)-based data association, followed by a Pareto-optimality-based position estimator and a residual-weighted velocity estimator. The framework is also extended to dual-polarized arrays. Simulation results demonstrate that the proposed cooperative scheme achieves better parameter estimation performance than conventional methods.

Unlike traditional parameter-based sensing methods, Huang \textit{et al.} introduced an image-based cooperative ISAC framework for low-altitude surveillance \cite{huang2025cooperative, huang2025learned}. The proposed approach formulates low-altitude sensing as a compressed-sensing (CS) reconstruction problem by leveraging multi-BS cooperation and the inherent sparsity of aerial images. To mitigate off-grid modeling errors, the authors develop a physics-embedded deep neural network that integrates physical priors with data-driven learning to refine image reconstruction. In addition, an online hard example mining (OHEM) strategy is incorporated into the loss function to enhance the detection of rare UAV targets in large low-altitude airspace. Simulation results demonstrate that this hybrid model-aided learning framework achieves a 97.55\% detection rate, significantly outperforming traditional CS-based methods under off-grid conditions.

\begin{table*}[!t]
\renewcommand{\arraystretch}{1.3}
\centering
\caption{Summary of UAV Detection and Tracking Using Networked-BSs ISAC.}
\label{tab:5}
\begin{threeparttable}

% 关键：m{} 列格式让内容垂直居中
\begin{tabular}{
    >{\centering\arraybackslash}m{3cm}|
    >{\centering\arraybackslash}m{2.5cm}|
    >{\centering\arraybackslash}m{5.3cm}|
    >{\centering\arraybackslash}m{5.0cm}
}
\toprule[1pt]
\textbf{Problems} & \textbf{References} & \textbf{Algorithms} & \textbf{Key Results} \\
\midrule

% =============================
% System Capability Modeling
% =============================
\multirow{2}{3cm}{%
    \centering
    \textbf{\makecell{System Design \\ and Architecture}}%
    \vspace{-5\baselineskip}%
}
& Chen \textit{et al.} \cite{chen2025iot} &
\begin{itemize}[leftmargin=*]
  \item IoT-enabled multi-BS cooperative sensing framework.
  \item Data-level and signal-level fusion.
  \item Spatial registration.
  \item Time alignment.
\end{itemize} &
\begin{itemize}[leftmargin=*]
  \item Achieves 85-90\% detection accuracy under optimal conditions.
  \item Reduces processing latency to 10-20 ms.
  \item Improves energy efficiency by 25-30\% via cooperative operation.
  \item Eliminates sensing blind zones and extends coverage across multi-BS network.
  \item Robust to urban clutter, multi-target, and adverse-weather scenarios.
\end{itemize}
\\ \cline{2-4}
& Feng  \textit{et al.} \cite{feng2025networked} &
\begin{itemize}[leftmargin=*]
  \item VSC construction with PBS/SBS cooperation.
  \item Multi-BS clutter suppression and 3D angle/range/Doppler estimation via MUSIC.
  \item Centralized EKF-based multi-BS fusion for 3D tracking.
  \item PBS handover and VSC inter-cell handover strategy.
\end{itemize}&
\begin{itemize}[leftmargin=*]
  \item Enables continuous multi-BS UAV tracking with seamless PBS/VSC handover.
  \item Significantly improves sensing robustness under blockage conditions.
  \item Achieves stable tracking of multiple UAVs.
\end{itemize}

\\
\midrule

\multirow{3}{3cm}{%
    \centering
    \textbf{\makecell{Parameter Estimation}}%
    \vspace{-15\baselineskip}%
}
& Lu \textit{et al.} \cite{lu2024integrated} &
\begin{itemize}[leftmargin=*]
  \item MUSIC-based angle/range/velocity estimation with spectral-superposition preprocessing to suppress noise and improve single-BS accuracy.
  \item Symbol-level fusion via lattice-point orthogonality search.
\end{itemize}&
\begin{itemize}[leftmargin=*]
  \item Large root mean square error (RMSE) reduction vs. single-BS and data-level fusion.
  \item Strong robustness across SNR variations.
  \item Most accurate localization and velocity among baselines.

\end{itemize}
\\ \cline{2-4}

& Zhang \textit{et al.} \cite{zhang2024target} &
\begin{itemize}[leftmargin=*]
  \item 2D-FFT OFDM delay/Doppler extraction.
  \item Bistatic range measurement and ML-based data association.
  \item Ill-conditioned measurement removal via geometric constraints.
\end{itemize}&
\begin{itemize}[leftmargin=*]
  \item Robust bistatic range accuracy under symbol timing offset (STO)/ carrier frequency offset (CFO).
  \item Accurate multi-target localization from communication-only signals.
  \item Improved robustness via cooperative AP fusion.

\end{itemize}
\\ \cline{2-4}

& Tang \textit{et al.} \cite{tang2025cooperative} &
\begin{itemize}[leftmargin=*]
  \item Spatial-smoothing tensor decomposition for joint AoA/range/Doppler/channel estimation.
  \item Reduced-dimensional GRQ-AoA estimator for reduced-complexity angle estimation.
  \item MST-based false-removing multi-BS data association.
  \item Pareto optimality-based position fusion and residual weighting-based velocity fusion.
\end{itemize}&
\begin{itemize}[leftmargin=*]
  \item Lower complexity than traditional methods.
  \item More robust association under false detections.
  \item Improved localization and true velocity accuracy through multi-BS fusion.

\end{itemize}

\\
\midrule

% =============================
% Beamforming & Trajectory
% =============================

\textbf{\makecell{Radio Image-based \\ Sensing}}
& Huang \textit{et al.} \cite{huang2025cooperative, huang2025learned} &
\begin{itemize}[leftmargin=*]
  \item CS-based cooperative radio image reconstruction using channel state information (CSI).
  \item Subspace pursuit for sparse on-grid imaging.
  \item Physics-embedded CNN imager with residual blocks and OHEM for accurate off-grid radio image reconstruction.
\end{itemize}&
\begin{itemize}[leftmargin=*]
  \item Achieves high-fidelity radio imaging with superior off-grid robustness.
  \item Achieves a 97.55\% detection rate.
\end{itemize}
\\

\bottomrule[1pt]
\end{tabular}

\end{threeparttable}
\end{table*}

\subsection{UAV Identification}
Target identification aims to determine the specific type or category of a detected object, such as distinguishing UAVs from birds or other low-altitude aerial targets. In ISAC-based UAV surveillance, it represents a key task following detection and tracking. Unlike traditional radar classification, ISAC-based identification is inherently coupled with the communication-centric sensing framework, where OFDM waveforms enable delay–Doppler analysis for micro-Doppler feature extraction, MIMO provides spatial discrimination, and multi-BS cooperation offers enhanced robustness through multi-perspective observations. However, reliable identification remains challenging due to limited sensing resources, strong clutter, and highly similar patterns among small aerial objects. Therefore, both signal processing and learning-based approaches have been explored for UAV identification. A summary of the related works in this subsection is given in Table \ref{tab:6}.

\begin{table*}[hbt]
\renewcommand{\arraystretch}{1.3}
\centering
\caption{Summary of Existing ISAC-based UAV Identification Methods.}
\label{tab:6}
\begin{threeparttable}

% 关键：m{} 列格式让内容垂直居中
\begin{tabular}{
    >{\centering\arraybackslash}m{3cm}|
    >{\centering\arraybackslash}m{2.5cm}|
    >{\centering\arraybackslash}m{4cm}|
    >{\centering\arraybackslash}m{3cm}|
    >{\centering\arraybackslash}m{4cm}
}
\toprule[1pt]
\textbf{References} & \textbf{Objectives} & \textbf{Algorithms} & \textbf{Dataset/Data Source} & \textbf{Key Results} \\
\midrule

Ma \textit{et al.} \cite{ma2024performance} & UAV identification using different TDD frame structures &
\begin{itemize}[leftmargin=*]
  \item \textbf{TDD frame design:} compares three TDD types for rotor micro-Doppler sensing.
  \item \textbf{Micro-motion feature extraction:} uses STFT for rotor spectrogram and parameter estimation.
\end{itemize} & Simulation based on MATLAB and FEKO software &
\begin{itemize}[leftmargin=*]
  \item Micro-Doppler features of rotor UAVs can be effectively extracted from 5G signals.
  \item Optimized TDD structure with denser downlink sensing slots achieves lowest rotor parameter errors under low SNR.
\end{itemize}
\\ \cline{1-5}

Wei \textit{et al.} \cite{wei2025uav} & UAV identification in clutter and interference &
\begin{itemize}[leftmargin=*]
  \item \textbf{ISAC TDD frame:} enables dense sensing and micro-Doppler extraction.
  \item \textbf{rmD-NSP:} separates weak rotor echoes from interference.
  \item \textbf{SET:} enhances time-frequency concentration of rotor features.
\end{itemize} & Simulation and Sub-6G ISAC hardware testbed
 &
\begin{itemize}[leftmargin=*]
  \item Captures eight rotations of the DJI M300 RTK UAV's rotor in urban environment within 0.1 s.
  \item Improves the integrity of the rotor micro-Doppler features by 60\%.
\end{itemize}

\\ \cline{1-5}

Luo \textit{et al.} \cite{luo2025pinpunet} & UAV type classification under micro-Doppler smearing &
\begin{itemize}[leftmargin=*]
  \item \textbf{PinpuNet:} employs inhomogeneous 2D convolutions to separately capture temporal and frequency correlations, while residual layers preserve subtle yet informative features across deep layers.
\end{itemize} & LSS-FMCWR-1.0 dataset
 &
\begin{itemize}[leftmargin=*]
  \item Achieves 97.17\% (4-class) and 99.50\% (5-class) classification accuracy on original data.
  \item Maintains above 91\% classification accuracy even under the smearing effect.
  \item Achieves at least 6.6\% higher accuracy than the second-best model.
\end{itemize}

\\ \cline{1-5}

Luo \textit{et al.} \cite{chu2025micro, luo2025airguard} & UAV and bird recognition &
\begin{itemize}[leftmargin=*]
  \item \textbf{AirGuard scheme:} extracts cmD spectrum and HRRP from ISAC echo signals.
  \item \textbf{Dual-feature fusion CNN:} jointly learns cmD and HRRP images for classification between UAV and bird.
\end{itemize} & Synthetic dataset generated from 3D mesh-based UAV and bird models
 &
\begin{itemize}[leftmargin=*]
  \item Achieves over 98\% classification accuracy when combining cmD and HRRP features.
\end{itemize}

\\ \cline{1-5}

Xue \textit{et al.} \cite{xue2025dc} & UAV and bird recognition &
\begin{itemize}[leftmargin=*]
  \item \textbf{DC-Former network:} integrates depth convolution for local spatial feature extraction and Transformer with multi-head self-attention for global dependency modeling.
\end{itemize} & Simulation and measured radar echo data using 5G NR signals at 3.5 GHz, 100 MHz bandwidth
 &
\begin{itemize}[leftmargin=*]
  \item Achieves 96-98\% classification accuracy.
  \item Outperforms conventional CNN-based models.
\end{itemize}

\\ \cline{1-5}

Costa \textit{et al.} \cite{costa2024modelling, costa2025modeling} & UAV micro-Doppler data generation for ISAC target classification &
\begin{itemize}[leftmargin=*]
  \item \textbf{OFDM-based bistatic model:} extends thin-wire formulation for 2D high-range resolution micro-Doppler data generation.
  \item \textbf{Multi-propeller extension:} supports independent rotor geometry and rotation speeds.
  \item \textbf{Body reflection modeling:} adds Gaussian/measurement-based RCS and vibration effects.
\end{itemize} & Simulation and BiRa bistatic measurement system
 &
\begin{itemize}[leftmargin=*]
  \item Simulated signatures closely match measured data.
  \item Enables large-scale synthetic data generation for UAV classification.
\end{itemize}

\\

\bottomrule[1pt]
\end{tabular}

\end{threeparttable}
\end{table*}

Micro-Doppler is a key feature for identifying UAVs. Ma \textit{et al.} evaluated UAV identification based on micro-Doppler features using 5G ISAC signals under different time-division duplex (TDD) frame structures \cite{ma2024performance}. By extracting rotor micro-motion spectrograms and estimating parameters such as rotation speed and blade length, the study shows that 5G signals can effectively capture micro-Doppler features of rotor UAVs. Simulation results show that a properly designed TDD frame structure can achieve better parameter estimation accuracy under low SNR conditions, underscoring the importance of optimizing TDD configurations for UAV identification in future mobile communication networks. Wei \textit{et al.} studied a UAV’s rotor micro-Doppler feature extraction for ISAC systems \cite{wei2025uav}. To address the difficulty of identifying low-mobility or hovering UAVs in cluttered environments, it designs a TDD frame structure configuration and introduces a rotor micro-Doppler null space pursuit (rmD-NSP) algorithm to isolate weak rotor-induced micro-Doppler components from body vibration and clutter. Combined with synchroextracting transform (SET) for high-resolution spectral enhancement, the method enables accurate extraction of UAV rotor signatures from CSI.

The discrimination between different UAV types and birds is also an important research direction. Luo \textit{et al.} investigated PinpuNet \cite{luo2025pinpunet}, a data-driven learning framework designed for classifying UAVs using smeared micro-Doppler spectrograms in ISAC systems. Radar echoes are first transformed into micro-Doppler spectrograms via short-time Fourier transform (STFT), which serve as inputs to the network. PinpuNet employs inhomogeneous 2D convolutions to separately capture temporal and frequency correlations, and integrates residual layers to preserve subtle yet informative features across deep layers, ensuring that fine-grained discriminative cues are retained during feature transformation. The proposed method is evaluated on the LSS-FMCWR-1.0 dataset \cite{xiaolong2023multiband}, which contains radar echoes of six UAV types including DJI Mavic 2, Phantom, Inspire 2, M350, M600, and a fixed-wing UAV, across K band (4-class) and L band (5-class) classification tasks. PinpuNet achieves classification accuracies of 97.17\% on the K band and 99.50\% on the L band, and maintains above 91\% accuracy under the smearing effect. Luo \textit{et al.} presented AirGuard, a dual-feature recognition framework for ISAC systems to effectively distinguish UAVs from birds \cite{chu2025micro, luo2025airguard}. The system extracts both the centralized micro-Doppler spectrum and the high-resolution range profile (HRRP) from echo signals, and employs a convolutional neural network (CNN) to fuse these features for target classification. Trained on synthesized samples, AirGuard demonstrates robust discrimination performance under noisy conditions, showcasing reliable UAV identification for ISAC surveillance. Xue \textit{et al.} proposed a depth convolution-former (DC-Former) network for UAV and bird recognition \cite{xue2025dc}. To address the limitations of simulated datasets, they collect real radar echo data of rotor UAVs and birds using 5G NR signals, and leverage STFT to generate spectrograms representing micro-Doppler signatures. A total of 10,000 time-frequency spectrograms for each class (UAV and bird) are constructed. The proposed DC-Former network takes a 3D time-frequency image as input and processes it through depth convolution and Transformer \cite{vaswani2017attention} modules to extract discriminative spatial-temporal features. The results show that the proposed method achieves a classification accuracy of 96–98\%, outperforming conventional CNN-based models. Furthermore, to address the lack of sufficient training data, Costa \textit{et al.} proposed a bistatic OFDM-based model to simulate UAV propeller micro-Doppler signatures in distributed ISAC systems \cite{costa2024modelling, costa2025modeling}. The simulated signatures exhibit a high correlation with measured data, validating the model's effectiveness and enabling large-scale synthetic data generation for UAV classification.

\subsection{Experimental Tests and Analysis}
Currently, most research on UAV surveillance using MIMO-OFDM ISAC systems relies primarily on simulation results to validate the theoretical methods. Meanwhile, several studies have also conducted experimental tests to demonstrate the preliminary feasibility of these approaches. This section reviews representative experimental efforts, with key test results summarized in Table \ref{tab:7}.

Yuan \textit{et al.} presented an experimental analysis and modeling of UAV RCS \cite{yuan2024experimental}. The measured RCS values range from approximately -40 dBsm to 5 dBsm and exhibit angle-dependent fluctuations. Three distributions are employed for data fitting, and the results show that the Rician distribution provides the best fitting performance. Moreover, under poor sensing conditions, the probabilities of the normalized RCS power falling below -20 dBsm and -10 dBsm are 1\% and 10\%, respectively, whereas under favorable conditions, the corresponding thresholds shift to -18.5 dBsm and -8.5 dBsm. These findings provide valuable insights into UAV RCS characterization for ISAC research. Gui \textit{et al.} implemented a prototype system and conducted field experiments on both ground vehicles and low-altitude UAVs \cite{gui2025mmw}. The system uses a dual-polarized 64-element antenna array operating at 25.75 GHz with a 95.04 MHz bandwidth. In the experiments, a UAV flying at approximately 10 m/s within a 100 m range is successfully detected and imaged using multi-frame ISAR reconstruction, producing a continuous image stream that clearly captures translational motion and enabled accurate velocity estimation. The results validate the feasibility of integrating ISAC with ISAR techniques for high-resolution sensing and surveillance of low-altitude UAVs. Liu \textit{et al.} proposed a circular fitting clutter suppression algorithm within a MIMO-OFDM ISAC framework for low-altitude UAV detection \cite{liu2025circular}. By reconstructing echo trajectories on the complex plane, the algorithm effectively suppresses static clutter and enhances UAV signal visibility in range-Doppler maps. Simulation results show that the proposed method improves the energy ratio by up to 41.85 dB for low-speed UAVs compared with the traditional moving target indicator (MTI) approach, significantly enhancing sensing performance in strong-clutter environments. The approach is further validated in Hangzhou, where UAV Remote ID data with missing segments are fused with 5G-A BS sensing results, allowing the sensing data to fill trajectory gaps and enhance continuity, thereby demonstrating robust clutter suppression and accurate UAV tracking in complex urban environments. Wei \textit{et al.} validated their proposed rotor micro-Doppler feature extraction technique on a 3.5 GHz Sub-6 GHz ISAC hardware testbed built on the 5G NR \cite{wei2025uav}. Using a DJI M300 real-time kinematic (RTK) UAV flying at 5 m/s, the BS successfully captures eight rotor rotations within the 0.1 s observation period, demonstrating accurate and robust rotor micro-Doppler extraction under real urban conditions.

Furthermore, a number of leading companies have carried out relevant field tests. ZTE Corporation and the State Key Laboratory of Mobile Network and Mobile Multimedia Technology designed and tested the world's first ISAC system powered by practical 5G networks \cite{liu2023integrated}. The system utilizes existing 5G BSs without any hardware modification to simultaneously detect, track, and localize UAVs, vehicles, and pedestrians while maintaining normal communication services. During simultaneous tracking experiments, a UAV flies at an altitude of approximately 30 m with a speed of 3 m/s across the test field, while a car crosses the area at a constant speed of 3 m/s and a pedestrian walks along the edge of the field. The localization errors for the three targets range from 0.3 m to 3.0 m. Further long-range tests demonstrate a maximum tracking distance of 1.43 km. These results obtained from the field trials confirm the feasibility of ISAC using 5G networks and pave the way for future research and engineering of practical ISAC systems. China Mobile Research Institute conducted outdoor UAV field experiments using an ISAC prototype operating in the mmWave band \cite{chih2025cooperative}. The prototype has a center frequency of 26 GHz, with both the sensing and communication bandwidths set to 200 MHz. Two synchronized BSs jointly detects a UAV hovering at an altitude of 500 m using standard OFDM waveforms. The system achieves sub-meter range estimation accuracy, demonstrating consistent performance in outdoor environments. These results verify that cooperative ISAC can realize high-precision UAV localization and demonstrate its feasibility for large-scale low-altitude surveillance in future 5G-A and 6G networks.

China Telecom Research Institute and Beijing University of Posts and Telecommunications (BUPT) validated a 3D ISAC sensing model through both an mmWave OTA anechoic chamber test and outdoor trials on a commercial mmWave ISAC network \cite{li2025low}. In the OTA setup, a 26 GHz ISAC BS and a radar target simulator are used to examine sensing boundaries and antenna FOV under controlled conditions. The OTA experiment achieves an average maximum sensing range of approximately 1000.22 m and an average minimum detectable distance of 0.04 m. The measured horizontal and vertical sensing angles are 59.95° and 40.28°, differing by merely 0.08\% and 0.7\% from their theoretical values, respectively, confirming the high accuracy and stability of the single-BS 3D sensing model. In the real-world scenario tests, a commercial 26 GHz ISAC network is deployed to evaluate practical sensing performance. Single BS UAV experiments report positioning accuracies at cumulative distribution function (CDF) 95\% of 5.9-7.5 m horizontally and 2.3-3.2 m vertically for horizontal trajectories, and 2.4-3.3 m horizontally and 2.3-5.0 m vertically for dedicated vertical flights, which align well with the theoretically predicted sensing range and blind-spot boundaries. In the multi-BS cellular-like topology, networked sensing tests at altitudes of 100 m and 300 m demonstrate horizontal and vertical accuracies of about 8.5-8.7 m and 4.5-4.8 m at CDF 95\%. These results demonstrate the feasibility, robustness, and scalability of the proposed 3D ISAC sensing model from ideal OTA conditions to practical networked deployments.

China Telecom Research Institute, BUPT and China Telecom Singapore Innovation Research Institute conducted a dual-band outdoor ISAC experiment at 3.5 GHz and 26 GHz for joint aerial and maritime target sensing \cite{yuan2025synergistic}. The trials can achieve joint estimation of range, velocity, and angle under practical clutter and interference conditions, and reconstructed trajectories closely match the global positioning system (GPS) data. The system attains sub-meter accuracy in UAV trajectory tracking at 26 GHz and a detection range of up to 1.6 km for vessels at 3.5 GHz. These results validate the effectiveness of high-low frequency collaborative sensing and support the advancement of full-domain ISAC technologies. China Mobile Chengdu Research and Development Institute and Huawei jointly conducted a UAV detection test at the Zigong Aviation Industrial Park \cite{su2024integrated}. The experiment employs a 26.9 GHz mmWave ISAC system and evaluated multiple flight trajectories of two mainstream DJI models, the Inspire 2 and Phantom 4. The results show that the proposed low-altitude UAV detection architecture achieves a target detection success rate of 91.2\%.

Overall, existing experimental studies on MIMO OFDM-enabled ISAC for UAV surveillance mainly validate the feasibility of detection and tracking, including parameter estimation, clutter suppression, and multi-frame trajectory reconstruction, while identification-related validation remains limited. Most experiments are conducted on single-BS platforms to verify core sensing and signal processing functions, whereas only a few recent works explore multi-BS or networked setups for improved localization accuracy and robustness. In terms of platforms, both sub-6 GHz and mmWave systems have been investigated, with sub-6 GHz testbeds focusing on basic sensing and micro-Doppler feature extraction, and mmWave platforms enabling high-resolution sensing and long-range detection, while recent dual-band experiments further demonstrate the potential of combining coverage and resolution. These results indicate that, although current experiments have confirmed the feasibility of ISAC-based UAV sensing, large-scale networked deployment and real-time identification remain important directions for future research.

\begin{table*}[hbt]
\renewcommand{\arraystretch}{1.20}
\centering
\caption{Representative Test Results for UAV Surveillance using MIMO-OFDM ISAC Systems.}
\label{tab:7}
\begin{threeparttable}

\begin{tabularx}{\textwidth}{
    >{\centering\arraybackslash}m{3.0cm}|
    >{\raggedright\arraybackslash}m{5.0cm}|
    >{\centering\arraybackslash}X |
    >{\centering\arraybackslash}X |
    >{\centering\arraybackslash}X |
    >{\centering\arraybackslash}X
}
\toprule[1pt]
\multirow{4}{=}{\centering \textbf{Institutes}} &
\multirow{4}{=}{\centering \textbf{Experimental Setup}} &
\multicolumn{4}{c}{\textbf{Results}} \tabularnewline
\cmidrule(lr){3-6}
& &
\textbf{Range/Angle Metric} &
\textbf{Localization Error} &
\textbf{Max Range} &
\textbf{Detection Rate} \tabularnewline
\midrule

\textbf{ZTE Corporation and State Key Laboratory of Mobile Network and Mobile Multimedia Technology} \cite{liu2023integrated} &
\textbf{BS configuration parameters:}
\begin{itemize}[leftmargin=*]
  \item Center frequency: 4.85 GHz.
  \item Sensing and communication bandwidth: 100 MHz.
  \item Transmit/Receive array: 64 elements.
\end{itemize} & -- &
\parbox[c]{\linewidth}{\centering 0.3--3.0 m\\(Simultaneous tracking of UAV, car, and
pedestrian)} &
\parbox[c]{\linewidth}{\centering 1.43 km\\(Long-range UAV tracking)} &
-- \\

\cline{1-6}

\textbf{China Mobile Research Institute} \cite{chih2025cooperative} &
\textbf{ISAC prototype parameters:}
\begin{itemize}[leftmargin=*]
  \item Center frequency: 26 GHz.
  \item Sensing and communication bandwidth: 200 MHz.
  \item Transmit/Receive array: 256 elements.
  \item EIRP: 58 dBm.
\end{itemize} &
  \parbox[c]{\linewidth}{\centering
Sub-meter level\\
(500 m altitude).}
 &
-- & -- & -- \\
\cline{1-6}

\multirow{8}{=}{\centering \textbf{China Telecom Research Institute \\ and BUPT} \\ \cite{li2025low}} &
\textbf{OTA anechoic chamber test:}
\begin{itemize}[leftmargin=*, nosep, topsep=0pt, after=\vspace{0.4em}]
  \item 26 GHz mmWave ISAC prototype system.
  \item Single-BS 3D sensing model (turntable-controlled AAU and target simulator).
\end{itemize} &
  \parbox[c]{\linewidth}{\centering Azimuth 0.08\%, elevation 0.7\% (Angle deviation).}
 &
-- &
  \parbox[c]{\linewidth}{\centering 1000.22 m}.
 &
-- \\
\cline{2-6}

& \textbf{Real-world scenario test:}
\begin{itemize}[leftmargin=*, nosep, topsep=0pt, after=\vspace{0.4em}]
  \item mmWave 5G-A commercial demonstration network.
  \item EIRP: 66 dBm (CW); 75 dBm (pulse).
  \item Urban areas/towns; DJI Mavic 3 UAV.
\end{itemize} &
-- &
  \parbox[c]{\linewidth}{\centering \textbf{Single-BS:} 5.9--7.5 m (H), 2.3--3.2 m (V) for horizontal flights; 2.4--3.3 m (H), 2.3--5.0 m (V) for vertical flights.
  \textbf{Multi-BS:} 8.5--8.7 m (H), 4.5--4.8 m (V).} &
-- &
-- \\
\cline{1-6}

\textbf{China Telecom Research Institute and BUPT and China Telecom Singapore Innovation Research Institute} \cite{yuan2025synergistic} &
\textbf{3.5 GHz system parameters:}
\begin{itemize}[leftmargin=*]
  \item Center frequency: 3.5 GHz; bandwidth: 100 MHz; EIRP: 46 dBm.
  \item Tx: 4 elements; Rx: 16 elements.
\end{itemize}
\textbf{26 GHz system parameters:}
\begin{itemize}[leftmargin=*]
  \item Center frequency: 26 GHz; bandwidth: 100 MHz; EIRP: 55 dBm.
  \item Tx/Rx: 256 elements.
\end{itemize} &
-- &
  \parbox[c]{\linewidth}{\centering 0.5--1.0 m}. &
  \parbox[c]{\linewidth}{\centering 220 m}. &
-- \\
\cline{1-6}

\textbf{China Mobile Chengdu Institute of Research and Development and Huawei} \cite{su2024integrated} &
\textbf{BS configuration parameters:}
\begin{itemize}[leftmargin=*]
  \item Center frequency: 26.9 GHz.
  \item Transmit power: 160 W.
  \item Horizontal beamwidth: 80°; vertical beamwidth: 20°.
\end{itemize} &
-- & -- & -- &
  \parbox[c]{\linewidth}{\centering 91.2\%}. \\
\bottomrule[1pt]
\end{tabularx}

\end{threeparttable}
\end{table*}

\section{Open Issues and Future Research Directions}
\label{sec4}
\begin{figure*}[hbt]
\centering
\includegraphics[scale=0.55]{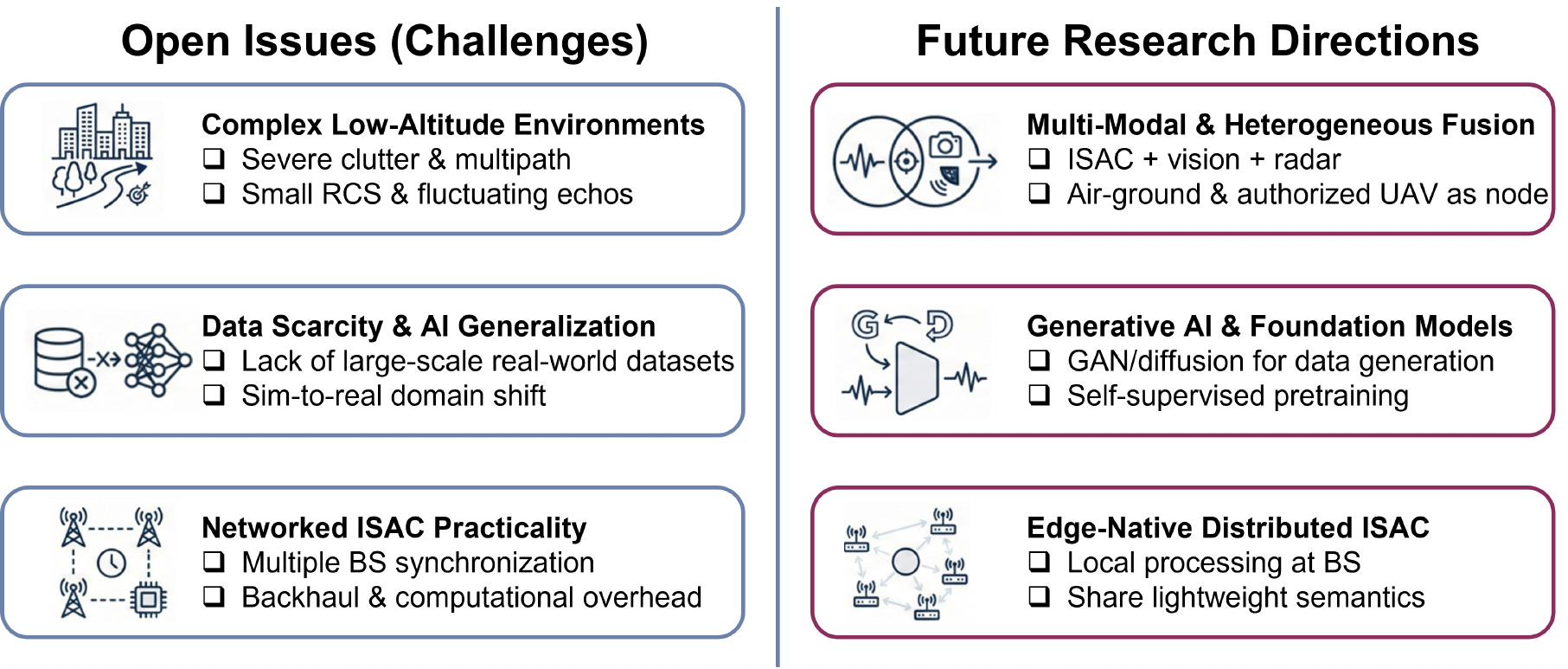}
\caption{Open issues and future research directions.}
\label{fig:7}
\end{figure*}

In this section, we first present the challenges of current UAV surveillance with MIMO OFDM ISAC and then conclude the future research trends, as illustrated in Fig. \ref{fig:7}.

\subsection{Challenges}
\subsubsection{Sensing Limitations in Complex Low-Altitude Environments}
Achieving reliable surveillance in low-altitude airspace is fundamentally constrained by the complexity of the propagation environment and the physical characteristics of small UAVs. Despite the advancements in MIMO-OFDM ISAC, several critical limitations remain:
\begin{itemize}
  \item
  \textbf{Severe Clutter and Multipath Interference:} The low-altitude channel, particularly in urban scenarios (e.g., urban canyons), imposes dual challenges on surveillance performance. First, strong environmental clutter poses a severe detection bottleneck. High-power echoes from non-target obstacles, such as static buildings and dynamic ground vehicles, can easily overwhelm the weak reflections of small UAVs. This clutter masking effect significantly reduces the SCNR, leading to high miss detection rates; for instance, simulations show up to 16\% miss detection at short ranges in dense urban environments. Secondly, multipath propagation further complicates sensing and tracking. Due to abundant reflectors like glass facades and metallic surfaces, part of the target's echo may return to the BS via indirect paths. These multipath components typically exhibit different delays and Doppler shifts, and are often misinterpreted as additional objects (ghost targets), thereby increasing false alarms and causing trajectory discontinuities.
  \item
  \textbf{Small RCS and Fluctuating Echoes:} Unlike traditional aircraft, consumer UAVs are typically made of plastic and carbon fiber, resulting in extremely small RCS. Experimental measurements indicate that UAV RCS values can range from approximately -40 dBsm to 5 dBsm \cite{yuan2024experimental}, exhibiting significant fluctuations depending on the observation angle. Under poor sensing conditions, the normalized RCS power frequently drops below detection thresholds (e.g., -20 dBsm), making consistent long-range detection challenging for standard cellular BS power levels. This necessitates high SNR requirements that are often difficult to meet at the cell edge.
\end{itemize}

\subsubsection{Data Scarcity and AI Generalization}
The success of learning-based detection and identification algorithms (e.g., CNNs, Transformers) heavily depends on the availability of high-quality, annotated training data. However, the development of such algorithms for ISAC-based UAV surveillance faces two intertwined hurdles:
\begin{itemize}
  \item
  \textbf{Scarcity of Large-Scale Real-World Datasets:} Unlike computer vision, there are no publicly available, large-scale, standard datasets for ISAC-based UAV sensing. Due to the high cost of experimental campaigns and strict flight regulations, the majority of existing research relies primarily on simulation results to validate theoretical methods. To mitigate this shortage, researchers often resort to synthetic data generation; for instance, the AirGuard framework is trained on synthesized samples, and recent works have focused on developing mathematical models to generate large-scale synthetic micro-Doppler signatures. While necessary, this reliance on synthesis creates a disconnect from physical reality.
  \item
  \textbf{The Sim-to-Real Generalization Gap:} Models trained on simulated data often suffer from poor generalization when deployed in real-world scenarios (i.e., the domain shift problem). Simulations typically employ idealized channel models that fail to fully capture complex environmental factors, such as the drastic angle-dependent fluctuations of UAV RCS or specific hardware impairments. Researchers explicitly highlight the limitations of simulated datasets, noting that real-world measurement data is essential for robust performance. Consequently, AI models that achieve high accuracy in simulations may exhibit significant performance degradation when facing the unmodeled noise, interference, and clutter of actual 5G/6G networks.
\end{itemize}

\subsubsection{Synchronization and Overhead in Networked ISAC}
To overcome the blind spots and limited coverage of single-BS systems, networked ISAC architectures are essential. However, realizing efficient multi-BS cooperation introduces significant implementation challenges:
\begin{itemize}
  \item
  \textbf{Stringent Synchronization Requirements:} Networked sensing imposes synchronization demands that far exceed those of standard communications. Effective data fusion requires precise spatial and temporal coordination among distributed nodes. For coherent processing techniques, phase and time alignment must be maintained at extremely high precision to prevent destructive interference during signal aggregation. Achieving this adaptive beam registration and cross-correlation timing across a wide-area network with varying propagation delays remains a complex engineering hurdle.
  \item
  \textbf{Excessive Backhaul and Computational Overhead:} There is a fundamental tradeoff between sensing accuracy and network overhead. High-performance centralized fusion engines require the transmission of raw sensing data or high-dimensional features to a central node, placing a heavy burden on the backhaul links. While distributed algorithms have been proposed to reduce backhaul overhead and hierarchical processing pipelines can lower latency, transmitting the necessary measurement data for real-time tracking still consumes significant bandwidth. This resource contention is particularly acute when the system must simultaneously support high-rate user communications and continuous UAV surveillance.
\end{itemize}

\subsection{Future Directions}
\subsubsection{Multi-Modal and Heterogeneous Network Fusion}
To overcome the intrinsic physical limitations of single-modality sensing, future surveillance systems must evolve toward holistic, multi-dimensional perception architectures. While MIMO OFDM ISAC offers robust all-weather operation, it lacks the semantic classification capabilities of optical sensors. One future research direction should focus on the deep fusion of ISAC radio data (e.g., range-Doppler maps) with visual or infrared imagery to cross-verify RF detections, thereby effectively distinguishing true UAVs from multipath ghosts and birds in complex urban environments. Furthermore, to eliminate blind spots in the spatial domain, the surveillance network must extend from 2D ground deployments to 3D air-ground cooperative topologies. Future research should explore protocols for authorized UAVs to act as flying ISAC nodes, fusing aerial viewpoints with ground data to achieve seamless 3D coverage and resolve occlusion issues caused by high-rise buildings.

\subsubsection{Generative AI and Foundation Models for RF Sensing}
The rapid evolution of AI provides a promising pathway to alleviate the data scarcity problem in ISAC-based surveillance. Current deep learning models depend heavily on limited, scenario-specific datasets, constraining their robustness in complex environments. Future work should explore generative AI methods, such as generative adversarial networks (GANs) and diffusion models, to synthesize high-fidelity RF data that capture realistic Doppler patterns, array responses, and multipath structures. Compared with fixed geometric channel models, generative approaches can learn richer statistical characteristics from limited measurements and produce diverse micro-Doppler and RF signatures at scale. In parallel, the field is shifting from task-specific architectures toward RF foundation models. Similar to large language models (LLMs), these models can be pre-trained on large amounts of unlabeled wireless signals and later fine-tuned for tasks such as UAV detection or identification with minimal labeled data, greatly enhancing generalization across different environments.

\subsubsection{Edge-Native Distributed ISAC Architecture}
To mitigate the stringent synchronization requirements and excessive backhaul overhead inherent in centralized networked ISAC, future surveillance architectures must transition toward edge-native distributed processing. While centralized fusion engines offer high-precision tracking, they impose unsustainable bandwidth demands and strict phase alignment constraints. Future research should prioritize sensing at the edge, where BSs perform local signal processing, such as clutter suppression and parameter estimation, and exchange only lightweight semantic information (e.g., target coordinates and velocity vectors) rather than raw samples. Furthermore, to ensure scalability across wide-area networks, the focus must shift from coherent to non-coherent fusion techniques. Future research should explore asynchronous data association algorithms that rely on geometric consistency rather than precise time synchronization, thereby enabling robust multi-BS cooperation with significantly reduced engineering complexity and latency.

\section{Conclusion}
\label{sec5}
This paper has presented a comprehensive survey on MIMO OFDM-enabled ISAC for low-altitude non-cooperative UAV surveillance. We first examine the fundamental signal characteristics of low-altitude UAV sensing, including strong ground clutter, rapid channel variations, and mixed near-/far-field propagation, and review corresponding waveform design and modeling approaches. Building upon this foundation, we summarize existing UAV surveillance techniques from four perspectives, namely system design and optimization, detection and tracking, identification, and experimental validation. Importantly, MIMO OFDM-enabled ISAC emerges as a practical core technology for low-altitude UAV surveillance, as OFDM waveforms enable delay-Doppler sensing within existing communication frameworks, MIMO provides high spatial resolution for target discrimination, and cellular infrastructure reuse supports wide-area and persistent monitoring with low deployment cost. Despite substantial progress, several challenges remain, including robust sensing under dense clutter and multipath, scalable multi-BS cooperative fusion, and reliable identification under limited labelled data. Future research is expected to focus on multi-modal fusion, AI-driven sensing, and distributed ISAC architectures, toward practical and scalable low-altitude surveillance systems.

\section*{Acknowledgments}
This work was supported by the Hong Kong Polytechnic University Start-up Fund under the Strategic Hiring Scheme titled ``Suppressing Multipath-Induced False Alarms for Robust 5G mmWave UAV Detection via Geometric Channel Modeling'' and the Research Centre for Low Altitude Economy (RCLAE), The Hong Kong Polytechnic University.

\bibliographystyle{IEEEtran}
\bibliography{papers}

\vfill

\end{document}